\documentclass[12pt,preprint]{aastex}

\usepackage{graphicx}
\usepackage{epstopdf}

\newcommand{\etal}{\mbox{et~al.}}

\def\deg      {{\ifmmode^\circ\else$^\circ$\fi}} 

\shorttitle{Evolution of the number density of disk galaxies to z~1}
\shortauthors{Sargent et al.}

\begin{document}

\title{The evolution of the number density of large disk galaxies\\ in COSMOS$^*$}

\author{ 
M.~T. Sargent\altaffilmark{1}, C.~M. Carollo\altaffilmark{1},
S.~J. Lilly\altaffilmark{1}, C. Scarlata\altaffilmark{1}, R. Feldmann\altaffilmark{1}, P. Kampczyk\altaffilmark{1}, A.~M. Koekemoer\altaffilmark{2}, N. Scoville\altaffilmark{3,4}, J.-P. Kneib\altaffilmark{5}, A. Leauthaud\altaffilmark{5}, R. Massey\altaffilmark{3}, J. Rhodes\altaffilmark{3,6}, L.~A.~M. Tasca\altaffilmark{5}, P. Capak\altaffilmark{3}, H.~J. McCracken\altaffilmark{7,8}, C. Porciani\altaffilmark{1}, A. Renzini\altaffilmark{9}, Y. Taniguchi\altaffilmark{10}, D.~J. Thompson\altaffilmark{11,12}, K. Sheth\altaffilmark{3,13}
}

\altaffiltext{1}{Department of Physics, ETH Zurich, CH-8093 Zurich, Switzerland}
\altaffiltext{2}{Space Telescope Science Institute, 3700 San Martin Drive, Baltimore, MD 21218, USA}
\altaffiltext{3}{California Institute of Technology, MC 105-24, 1200 East California Boulevard, Pasadena, CA 91125, USA}
\altaffiltext{4}{Visiting Astronomer, Univ. Hawaii, 2680 Woodlawn Dr., Honolulu, HI, 96822, USA}
\altaffiltext{5}{Laboratoire d'Astrophysique de Marseille, BP 8, Traverse du Siphon, 13376 Marseille Cedex 12, France}
\altaffiltext{6}{Jet Propulsion Laboratory, 4800 Oak Grove Drive, Pasadena, CA 91109, USA}
\altaffiltext{7}{Institut d'Astrophysique de Paris, UMR 7095 CNRS, Universit\'e Pierre et Marie Curie, 98 bis Boulevard Arago, 75014 Paris, France}
\altaffiltext{8}{Observatoire de Paris, LERMA, 61 Avenue de l'Observatoire, 75014 Paris, France}
\altaffiltext{9}{Dipartimento di Astronomia, Universit\`a di Padova, Vicolo dell'Osservatorio 2, I-35122 Padua, Italy}
\altaffiltext{10}{Physics Department, Graduate School of Science, Ehime University, 2-5 Bunkyou, Matuyama, 790-8577, Japan}
\altaffiltext{11}{Caltech Optical Observatories, MS 320-47, California Institute of Technology, Pasadena, CA 91125, USA}
\altaffiltext{12}{Large Binocular Telescope Observatory, University of Arizona, 933 N. Cherry Ave., Tucson, AZ 85721, USA}
\altaffiltext{13}{Spitzer Science Center, California Institute of Technology, Pasadena, CA 91125, USA}

\altaffiltext{$\star$}{Based on observations with the NASA/ESA {\em Hubble
Space Telescope}, obtained at the Space Telescope Science Institute, which
is operated by AURA Inc, under NASA contract NAS 5-26555; also based on data
collected at: the Subaru Telescope, which is operated by the National
Astronomical Observatory of Japan; the XMM-Newton, an ESA science mission
with instruments and contributions directly funded by ESA Member States and
NASA; the European Southern Observatory, Chile; Kitt Peak National
Observatory, Cerro Tololo Inter-American Observatory, and the National
Optical Astronomy Observatory, which are operated by the Association of
Universities for Research in Astronomy, Inc. (AURA) under cooperative
agreement with the National Science Foundation; the National Radio Astronomy
Observatory which is a facility of the National Science Foundation operated
under cooperative agreement by Associated Universities, Inc.; and the
Canada-France-Hawaii Telescope operated by the National Research Council (NRC) of
Canada, the Centre National de la Recherche Scientifique de France and the
University of Hawaii.}

\begin{abstract}
We study a sample of approximately $16,500$ galaxies with $I_{ACS,AB}\leq22.5$ in the central 38\% of the COSMOS field (\cite{scov06}, this volume),
which are extracted from a catalog constructed from the Cycle 12 ACS F814W COSMOS data set.
Structural information on the galaxies is derived by fitting single S\'ersic (\cite{sers68}) models to their two-dimensional surface brightness
distributions. \\In this paper we focus on the disk galaxy population (as classified by the Zurich Estimator of Structural Types, ZEST;
\cite{scar06}, this volume), and investigate the evolution of the number density of disk galaxies
larger than approximately 5 kpc between redshift $z\sim1$ and the present epoch. Specifically, we use the measurements of the half-light radii derived from the S\'ersic fits to construct, as a
function of redshift, the size function $\Phi(r_{1/2},z)$ of both the total disk galaxy population and of  disk galaxies split in four bins of
bulge-to-disk ratio. In each redshift bin, the size function specifies the number of galaxies per unit comoving volume and per unit half-light radius
$r_{1/2}$. Furthermore, we use a selected sample of roughly 1800 SDSS galaxies to calibrate our results with respect to the local universe. We find that:\\ 
$(i)$ The number density of disk galaxies with intermediate sizes ($r_{1/2}\sim$ 5-7 kpc) remains nearly constant from $z\sim1$ to today. Unless the
growth and destruction of such systems exactly balanced in the last eight billion years, they must have neither grown nor been destroyed over this period.\\
$(ii)$ The number density of the largest disks ($r_{1/2}>7$ kpc) decreases by a factor of about two out to $z\sim1$.\\
$(iii)$ There is  a constancy - or even slight increase - in the number density of large bulgeless disks out to $z\sim1$; the deficit of large disks
at early epochs seems to arise from a smaller number of {\it bulged} disks. \\ Our results indicate that the bulk of the large disk galaxy population
has completed its growth by $z\sim1$, and support that secular evolution processes produce  -- or at least add stellar mass to -- the bulge components
of disk galaxies.
\end{abstract}


\section{Introduction}

A detailed quantification of galaxy structure as a function of redshift is an essential tool to study the evolution of galaxies through
cosmic time.  In particular, the evolution of galaxy sizes is a direct diagnostic of their star formation and mass assembly history. For
disk galaxies, several theoretical predictions have been made, which start being testable with observations:  For example, under the
assumption that the scale lengths of disks scale with their virial radii, their sizes are predicted to evolve as $r_{s}\propto H(z)^{-2/3}$ at
a fixed halo mass (\citet{fallefstathiou80}, \citet{mao98});  \cite{bouwenssilk02}, instead, assume a dependence of the star formation rate on the
local gas density within the disk, and infall of metal-free gas, and predict the sizes of disks to evolve as $R(z)/R(0) = 1 -
0.27\cdot z$.

Pioneering observations, however, generally appear to support a substantially weaker size evolution of the disk galaxy population than
predicted by theory. Data from the Canada-France Redshift Survey (CFRS) for a few tens of large disk galaxies up to redshift $z\sim1$  indicate  about
a magnitude of surface brightness  evolution for the disks, and a constant number density of large disks over this period (\cite{schade95};
\cite{schade96}; \cite{lil98}). Similar results are reported by \cite{roche98}. In contrast, 
\cite{simard99} interpret their data as indicating no surface brightness evolution in disks over the same redshift range. The work of
\cite{trujillo04}, on the other hand, provides further evidence for surface brightness evolution in galaxy disks, although a possible decrease in the sizes of
disks by approximately 30\% at $z\sim$ 0.7 is also proposed as an alternative to explain the data. Using 1500 disk galaxies at redshifts 0.25 $<z<$
1.25 from the GOODS survey, \cite{ravindranath04}  find at most a  very modest decrease of the number of disks with half-light radius $r_{1/2}>$ 4
kpc. The GEMS survey has been used to investigate the evolution with redshift of the  luminosity-size relation for disk galaxies (\cite{barden05}; see
also \cite{trujillo05}), but  no analysis of the evolution of the number density of disk galaxies as a function of galaxy size has been presented
based on the GEMS data.  Therefore, the observational evidence for a relatively constant number of (large) disks since $z\sim 1$ relies on very small
samples extracted from small areas in the sky.

In this article we use the COSMOS survey (\cite{scov06}, this volume) to  investigate how the number density of disk galaxies with sizes above 5 kpc -
 the threshold above which our sample is complete - evolves with cosmic time from $z\sim1$ to $z=0$. Thanks to the large statistics of COSMOS, we are
 able to extend the investigation so as to study this evolution as a function of the bulge-to-disk ratio of disk galaxies. Specifically, we use the
 ACS COSMOS database, and select a $I_{ACS,AB}\leq22.5$ sample from the 38\% of the COSMOS area that was imaged by the HST during Cycle 12. The
 complete $I_{ACS,AB}\leq22.5$ sample contains more than 16,000 galaxies.
We perform two-dimensional GIM2D (\cite{marlsimard98}) fits to their surface brightness distributions, which we describe with a single S\'ersic law
(\cite{sers68}). Free parameters in the fits are: the total flux $F_{tot}$, the S\'ersic index $n$, the half-light radius $r_{1/2}$, the ellipticity
$\epsilon = 1-b/a$ (with $a$ and $b$ the semimajor- and minor-axis), and the position angle $\phi$.

We compare the structural parameters obtained with the GIM2D  surface brightness fits with the structural measurements derived using  the
Zurich Estimator of Structural Types (ZEST; \cite{scar06}, this volume). ZEST uses individual measurements of four non-parametric
diagnostics of galaxy structure, namely, the concentration $C$, the asymmetry $A$, the Gini coefficient $G$, and the second order moment of
the brightest 20\% of the galaxy pixels, $M_{20}$ (c.f. \cite{scar06}, this volume, and references therein), plus the elongation of the
distribution of the light, to classify galaxies in three morphological types: early-types, disk galaxies or irregular and peculiar galaxies.
The disk galaxies are further split by ZEST into four classes of bulge-to-disk ratio ($B/D$). At any redshift, the multi-dimensional ZEST
classification grid substantially reduces the intermixture of structurally different galaxy populations compared with simpler approaches.
The comparison of our parametric structural quantities with the diagnostics supplied by ZEST shows the expected correlations (e.g., between
S\'ersic index $n$ and concentration $C$) and, by doing so, illustrates the robustness of the results of our GIM2D fits. It also demonstrates that
studies of selected galaxy populations which are classified by a simple cut in S\'ersic index $n$ - as customarily done, for instance, in studies of
the SDSS galaxy population and in other surveys of the high-$z$ universe - are affected by  mixing of galaxies with different structural properties.

We use the ZEST classification to extract, from the total  $I_{ACS,AB}\leq22.5$ sample, the subsample of approximately 12,000 disk galaxies (with
different $B/D$ ratios), and   use the half-light radii provided by our S\'ersic fits to compute  the number of disk galaxies per unit comoving volume
and per unit half-light radius $r_{1/2}$ as a function of redshift (the {\it size function} $\Phi(r_{1/2},z)$).  In order to compare the results  at
higher redshift with the local universe, we use a suitably selected sample of  about 1800 SDSS galaxies, which have been ``redshifted" to $z=0.7$ by
properly re-scaling for flux dimming, distance and image resolution (\cite{kampczyk06}, in preparation).

We find that the number density of disk galaxies with intermediate sizes $r_{1/2}\sim$ 5-7 kpc remains nearly constant from $z\sim1$ to today, while
that of the largest disks ($r_{1/2}>7$ kpc) shows a drop by a factor of approximately two. This deficit appears to stem mostly from a decline in the
number density of {\it bulged} disks at early epochs.\footnote{In the rest of the paper, we consistently use AB magnitudes (\cite{oke74}), and  thus
drop the subscript ``AB" from our notation. We adopt $\Omega_{m}=0.25$, $\Omega_{\Lambda}+\Omega_{m}=1$ and a present day Hubble parameter of $H_0=70~km~s^{-1}~Mpc^{-1}$.}

\section{Data analysis}

\subsection{The initial catalog: 16,538 ACS-selected galaxies}

We use a sample of 16,538 COSMOS galaxies with $I\leq22.5$, detected in the 259 ACS F814W images acquired during the HST Cycle 12
observing period (\cite{scov06}). Details on the processing of the ACS data are given in \cite{koek06}; a discussion of the ACS-detected catalog used as the starting point to build our sample is presented in \cite{leauthaud06}. In the following we give a brief overview over the steps leading to the production of the catalog:

\begin{itemize}

\item The ACS-detected catalog was generated by running SExtractor \citep[version 2.4.3]{bert96} twice on the reduced ACS images; the first time in a ``cold" run with a configuration optimized for the detection of large, bright objects, and then in a ``hot" run with a configuration optimized for small and faint sources.  
 
\item The two resulting samples were then merged together to produce a final catalog by retaining all the ``cold" detections plus the ``hot'' detections which fell outside the SExtractor segmentation map of any galaxy detected in the ``cold" run. This ACS-based catalog contains approximately 55,700 galaxies down to $I=24$.

\item Stars were removed by deleting the sources with SExtractor CLASS\_STAR$>0.6$ from the catalog.

\item The ACS-based catalog was inspected to remove all remaining over-deblended large galaxies as well as false detections from it. About 4.5\% of the sources in the original ACS-based catalog were deleted from our final catalog after this visual check.

\end{itemize}

The final and cleaned ACS-based catalog contains 55,651 galaxies down to $I=24$, 16,538 of which have $I\in[16,22.5]$ and an ACS-stellarity
parameter smaller than 0.6\footnote{Note that by applying the customary cut at a value of the stellarity parameter of 0.9 some very bright stars would be included in the sample. We inspected the sources with stellarity larger than 0.6 to ensure that no galaxy was excluded from our sample.}. These roughly 16,500 galaxies constitute the sample that we study in this article.

\subsection{Two-dimensional GIM2D fits}
 
The GIM2D (Galaxy IMage 2D) IRAF software package is designed for the quantitative structural analysis of distant galaxies (\cite{marlsimard98}, \cite{simard02}). GIM2D uses the Metropolis algorithm to converge to the analytical model, i.e. it carries out a Monte Carlo sampling of the likelihood function $P(w|D,M)$, which measures the probability that the parameter combination $w$ is the correct one given the data $D$ and the model $M$. The Monte Carlo approach of sampling the complex multidimensional topology of parameter space has the virtue of converging consistently to the same best fit model for a wide range of initial guesses. 

In this analysis, we adopt a simple single S\'ersic (\cite{sers68}) profile to describe the two-dimensional surface brightness distribution of a galaxy's light:
\begin{equation}
\Sigma(r)=\Sigma(R_{1/2})~exp\left\{-k\left[\left(\frac{r}{R_{1/2}}\right)^{1/n}-1\right]\right\}~,
\end{equation}
where the value of $k$ is chosen such that $R_{1/2}$ represents the radius containing half of the total flux. (Throughout this article we adopt the convention that $R_{1/2}$ stands for the apparent half-light radius on the image, measured in pixels, while $r_{1/2}$ denotes the physical half-light radius in units of kpc). The flexible form of the S\'ersic law has the advantage of parametrizing, through the variable exponent $n$, surface brightness distributions including the exponential radial fall-off of the light profile in bulgeless disks ($n=1$), and the classical de Vaucouleur profile encountered in elliptical galaxies ($n=4$).
  
In the case of a single S\'ersic model, GIM2D seeks the best fitting values for the following eight parameters: the total flux $F_{tot}$ (integrated to $r=\infty$); the half-light radius $R_{1/2}$; the ellipticity $e=1-b/a$, where $a$ and $b$ are the semimajor and semiminor axes of the brightness distribution; the position angle $\phi$; the offsets $dx$ and $dy$ from the initially specified centre of the galaxy; the residual background level $db$ and the value of the S\'ersic index $n$.

The procedure followed to obtain S\'ersic models for all 16,538 galaxies in our sample involved the steps listed below:
\begin{itemize}

\item \underline{Definition of the area to fit.} For each galaxy, GIM2D was
  instructed (by the means of a \textit{mask image}) to fit the distribution
  of light within an elliptical area with a semimajor axis equal to 1.5
  Petrosian radii (\cite{petro76})\footnote{The Petrosian radius $r_P$ is the
    radius at which the surface brightness in a thin annulus equals a given
    fraction of the average surface brightness within that radius, in our case
    a fraction of 20\%. Since it is computed from the ratio of two surface
    brightness values, the Petrosian radius is not as sensitive as other size
    measures to the effect noise has on objects at higher redshifts. It thus
    provides a robust and consistent definition of the surface over which to
    perform our GIM2D fits.} of the target galaxy and an ellipticity equal to
  the flattening of the SExtractor segmentation map (the latter defines the
  pixels associated with the object). The orientation of this Petrosian ellipse is also provided by SExtractor. During the
  fitting, the algorithm considers both pixels assigned to the objects of interest and those flagged as background in the {\it mask image}.
  The goodness of the fit (i.e. the $\chi^2$), on the other hand, is estimated based on the difference betweeen image and model for pixels within the Petrosian
  ellipse.

\item \underline{Generation of stamp images.} For each galaxy we extracted from the reduced ACS images, a {\it stamp image}, centered on the coordinates of the ACS-based catalog. The {\it stamp images} were $(i)$ no smaller than $10''\times10''$, and $(ii)$ always sufficiently large to guarantee a proper computation of the local background for all target galaxies. Point $(ii)$ is ensured by choosing the stamp size such as to contain at least the same number of ``background" pixels outside the masking area, as lie within an ellipse with semimajor axis equal to a single Petrosian radius, centered on the target galaxy.
 
\item \underline{Generation of cleaned images.} To optimize the quality of the fits, we allowed GIM2D to improve the determination of the center of the galaxy with respect to the coordinates specified in the catalog. Furthermore, GIM2D requires and computes the value of the background to produce the fit to the galaxy light.  Bright objects that are located near the galaxy to be fit can cause an inaccurate calculation of the background, and  an incorrect estimate of the galaxy center. To prevent these potential errors from affecting the quality of our S\'ersic models, we cleaned the galaxy stamp images of all close companions. In this cleaning procedure pixels in the stamp image that were associated - based on a segmentation map accompanying the ACS-catalog - with sources other than the one of interest were replaced by values reproducing the properties of the surrounding background (c.f. \cite{scar06}, this volume, for details). We used these {\it cleaned images} in the next steps.

\item \underline{Symmetrization of galaxy images.} We used GIM2D to produce a symmetrized version of the cleaned galaxy image. This was done by rotating the galaxy by $180^{\circ}$, subtracting the rotated image from the cleaned image, setting all pixels with values smaller than $2\sigma$ of the {\it cleaned image}'s background in the difference image to zero, and, finally, subtracting this clipped difference image from the {\it cleaned image}. These {\it symmetrized images} were used as input to GIM2D in order to converge to a satisfactory model of the galaxy light distribution more reliably and quickly. 

\item \underline{Initial launch parameters and allowed parameter space.} Given the robustness of GIM2D with respect to changes in the initial parameters (see Appendix A.1), we started all our fits with the same initial guesses of galaxy structural parameters, namely: $I=21$, $r_{1/2}=10~px$, $\epsilon=0.5$, $\phi=0^{\circ}$ and $n=4$. The calculation of a pure S\'ersic profile is achieved by fixing the ratio of flux in the bulge to the total flux, $B/T$, to unity. Structural parameters were allowed to vary within the following ranges: $I\in[17.5,26.7]$, $r_{1/2}\in[1~px,50~px]$, $\epsilon\in[0,1]$, $\phi\in[-180^{\circ},180^{\circ}]$, $n\in[0.2,9]$. We inspected the relatively small number of galaxies that appeared to genuinely require values outside these ranges, and allowed the parameters to converge to best fits outside those boundaries when our visual inspection confirmed the quality of the fits.

\item \underline{Adopted Point Spread Functions (PSFs).} We select a different PSF for each individual object based on its position on the tile and the current focal length of the ACS due to in-orbit breathing at the time of the exposure (\cite{rhod06}, this volume).

\end{itemize}

\subsection{Deletion of sources from the original ACS-based catalog}

A clear assessment of the selection effects affecting galaxy samples is essential to perform statistical studies of the evolution of galaxy populations. Sources had to be removed from our initial sample of 16,538 ACS-selected, Cycle 12 COSMOS galaxies with $I\in[16,22.5]$ and ACS-stellarity smaller than 0.6 due to a variety of reasons:

 \begin{itemize}
 
 \item A total of 465 sources (i.e., about 2.8\% of the sample) were deleted from our final sample, due to the failure of GIM2D to converge. 
 
 \item 29 additional sources were also excluded on the basis of the $\chi^2$ of the GIM2D fits. After inspection of the GIM2D residuals and of the $\chi_{GIM2D}^2$ distribution, we fixed the threshold for deleting sources from our sample to $\chi_{GIM2D}^2=15$. While this may seem an exceedingly generous criterion, visual inspection of all GIM2D fits and their residuals for objects with $\chi_{GIM2D}^2>1.5$ showed that, in the vast majority of the cases, GIM2D produces a fair model of the diffuse galaxy light, and the high values of $\chi_{GIM2D}^2$ in these objects were produced by structure such as knots of star formation, spiral arms or bars.
Figure 1 and the first three entries in Table 1 display and list three random examples of objects with $\chi_{GIM2D}^2 \in [5,15]$.  It is clear from the figure that the presence of spiral and other structure can lead to high values of $\chi_{GIM2D}^2$ even if the underlying symmetric component is reproduced well by the S\'ersic model. 

\item Of the galaxies in our sample, 344 lacked ground-based photometric information in at least five of eight bands; therefore, no accurate  photometric redshifts could be derived for these sources. 

\item A fourth selection criterion for the rejection of objects was applied in part of the analysis, based on the value of $\chi_{phz}^2$
supplied by ZEBRA (c.f. Section 4.1). As a test to evaluate the impact of high values of $\chi_{phz}^2$ in the photometric redshift
estimates, we adopted the approach of calculating the size functions (Section 4)  both by applying no $\chi_{phz}^2$ cut, and by deleting
from our sample sources with a $\chi_{phz}^2$ above a threshold given by the value corresponding to two standard deviations  in an ideal
$\chi^2$-distribution of a varying number of degrees of freedom (which in our case are the number of filters for which ground-based
photometric measurements are available). This cut  - when applied - reduces the sample by 1369 objects. 

\item Finally, objects with one ground-based set of coordinates but two distinct coordinates in the ACS-selected catalog must be expected to have ambiguous photometric information and they were therefore removed as well. A total of 271 objects fell in this category.

\end{itemize}

\section{The ZEST morphological classification of our COSMOS sample}

All COSMOS galaxies in our sample have been classified with ZEST, the Zurich Estimator of Structural Types, discussed in \cite{scar06} in this volume. ZEST is a multi-dimensional classification grid that simultaneously uses: $(i)$ the non-parametric diagnostics of asymmetry $A$, concentration $C$, Gini coefficient $G$, and the second order moment of the brightest 20\% of galaxy pixels M$_{20}$, (e.g., \cite{lotz04}), plus $(ii)$ the elongation of the galaxy's distribution of light, to classify galaxy types. 
For each galaxy, the final ZEST classification provides quantitative information on galaxy structure, and a classification into either of the types 1, 2 or 3 for early-type, disk and irregular or peculiar galaxies, respectively. ZEST further splits the disk galaxies into four sub-classes of bulge-to-disk ratio, ranging from bulgeless disks (type 2.3) to bulge-dominated spirals (type 2.0). The S\'ersic indices presented in this paper were employed in ZEST to check and refine the four $B/D$ subclasses of disk galaxies.

Figures 2 to 7 show, for the different ZEST types and for disk galaxies with different $B/D$ ratios, a few examples of GIM2D fits picked
randomly among representatives of the morphological class in question. In each figure we plot (from left to right) the original stamp image, the cleaned image, the GIM2D model and the fractional residuals between the observed galaxy and the analytical model. It is clear from these examples that early-type and disk galaxies are well described by the analytical fits with the S\'ersic profile. They do well in reproducing the smooth component underlying the galactic light and, as is visible in the residuals, leave behind only the sub-structure within the galaxies, such as spiral arms and bars. On the other hand, the GIM2D fits for the irregular galaxies are understandably less reliable, as the smooth and symmetric analytical models cannot capture the richness of structure of this class of objects.

To further highlight the robustness of the GIM2D fits, we show in Figure 8 the comparison of our values of the S\'ersic index with the four non-parametric diagnostics of galaxy structure computed to derive the ZEST classification of the galaxies in our sample. The expected correlations between diagnostics such as concentration and S\'ersic index, as well as the tightness of these relationships, are an independent confirmation of the reliability of the analytical fits.

The left hand side of Figure 9 shows the distribution of S\'ersic indices $n$ as a function of ZEST galaxy type. The figure illustrates that, as expected, the bulk of late-type galaxies (including the type 3 irregular galaxies) typically shows low values of $n$ and preferentially lies close to the typical value $n=1$ of disk-like exponential profiles. The distribution of early-type galaxies, on the other hand, reaches its peak at  $n\sim3.5$. The right panel of the figure shows the distribution of $n$ for disk galaxies of different $B/D$ ratio; by construction, the mode shifts to ever higher values of $n$ as the bulge component gains in importance. It is evident from Figure 9 that a separation of galaxy types according to a simple cut in S\'ersic index brings about a significant mixture of different galaxy populations. Depending on the scientific goals, however, it may be important to use samples as uncontaminated as possible by other galaxy types. In the following we thus adopt the more precise ZEST classification of galaxy types - rather than a cut in S\'ersic index - to investigate the evolution of the size function of disk galaxies (Section 4.3).

As a final illustration of the COSMOS sample under investigation, we show the distributions of physical sizes $r_{1/2}$ in kpc in Figure 10.
ZEST types and sub-types are displayed with different line styles, as in Figure 9. The distribution of half-light radii of disk-dominated galaxies (morphological type 2) peaks at roughly 2.75 kpc, that of elliptical galaxies (type 1) around 2 kpc. In the right panel of the figure, the peaks of the distribution of half-light radii shift to progressively smaller sizes with increasing $B/D$-ratio, as is expected for objects with successively more centrally concentrated surface brightness profiles. The only exception to this trend are the most bulge-dominated sources among the disk galaxy population (type 2.0), which break out of this sequence. The ZEST class 2.0 is defined as containing objects with a spheroidal component (and S\'ersic index) similar to the type 1 early-type galaxies, but, as opposed to the latter, showing an obvious disk component as well. For many scientific applications, it might prove beneficial to merge this morphological class with the type 1 early-type galaxies (c.f., for instance, \cite{scar06b} for a discussion on the evolution of early-type galaxies in COSMOS, in which the effects of including the ZEST type 2.0 class in the sample is discussed in detail). Here we emphasize that the $I\leq 22.5$ COSMOS sample under scrutiny encompasses, for each galaxy type individually, a broad range in physical sizes and allows for statistically sound statements thanks to the large number of available objects. Thus, with proper care given to the effects of incompleteness, this permits us to study the evolution with cosmic time of the comoving number density of galaxies of a given size. 

In this analysis we focus exclusively on the disk galaxies - classified as such by ZEST -  and postpone the investigation of other galaxy populations to future publications. Of the 14,520 galaxies reliably fitted with GIM2D that remain in the sample after applying the selection criteria in magnitude, stellarity parameter, $\chi_{GIM2D}^2$,  $\chi_{phz}^2$ and photometric reliability (see Section 2.3), 11,744 galaxies are classified as disk galaxies on the ZEST COSMOS classification grid. If no cut based on the value of $\chi_{phz}^2$ is applied, the remaining sample of disk galaxies contains 12,701 objects. Figure 11 shows the distribution with redshift of the galaxies in our sample and Tables 2 and 3 summarize in more detail how the disks and other galaxy types are assigned to the different investigated redshift bins up to $z\sim1$. This information is provided both for the case when a cut based on the value of $\chi_{phz}^2$ is applied to the sample, and the case when no such cut is applied. We have tested that the results obtained with and without the cut in $\chi_{phz}^2$ show no significant difference. We therefore show our results using the sample \textit{with} the cut in $\chi_{phz}^2$. Doing so implies the exclusion of a total of  $8\%$ of the original ACS-based $I\leq22.5$ Cycle 12 COSMOS sample.

\section{The Evolution of the  Size  Function of disk galaxies in COSMOS}

\subsection{ZEBRA photometric redshifts}

Several codes and approaches have been used to derive accurate photometric redshifts for COSMOS galaxies down to faint magnitudes (c.f. \cite{mobasher06} in this volume for a review). In our work, we adopt the photometric redshift estimates obtained by \cite{feld06} for the COSMOS galaxies of our ACS-selected $I\leq 22.5$ galaxy sample. These are computed with the Zurich Extragalactic Bayesian Redshift Analyzer (ZEBRA).

ZEBRA produces two separate estimates of the photometric redshifts of individual galaxies: A maximum likelihood estimate and a fully two-dimensional Bayesian estimate in the space of redshift and templates. In both approaches, ZEBRA uses an iterative technique to automatically correct the original set of galaxy templates to best represent the spectral energy distributions of the galaxies in different redshift bins. The availability of a ``training set" of spectroscopically derived zCOSMOS redshifts (c.f. \cite{lil06}, this volume) for a small fraction of the whole considered photometric sample allows for a precise calibration of the ZEBRA photometric redshifts and thus for an optimal correction of the galaxy templates.

The ZEBRA uses the available ancillary ground-based multi-band photometry to estimate the photometric redshifts of the COSMOS galaxies. $B$, $V$, $g$, $r'$, $i'$, and $z'$ fluxes are measured with Subaru (\cite{tani06}), $u^*$ fluxes with CFHT and the 4 m telescopes at Kitt-Peak and CTIO provide the $K_s$ photometry (\cite{capak06}). 

In the following, we use the ZEBRA maximum likelihood photometric redshifts of \cite{feld06} for our discussion of the evolution of the size function of disk galaxies. The maximum likelihood ZEBRA photometric redshifts have, at all redshifts of interest, an accuracy of $\Delta z/(1+z)\sim 0.03$ if compared with the zCOSMOS spectroscopic redshifts (\cite{lil06}). 
The application of ZEBRA to our sample shows some dependence of the photometric redshifts on whether small systematic offsets (of order $~0.05$ magnitudes or smaller), which are detected by the code, are applied to the calibration of the Subaru data. This has, however, no substantial impact on our analysis, as we show in the following by using, when relevant, the two ZEBRA estimates of the photometric redshifts obtained with and without corrections for these photometric offsets. The inclusion or exclusion of bad ZEBRA fits from our final sample does not change the conclusions of this manuscript.

\subsection{Derivation of the Size Function and assessment of completeness}

In each redshift bin, the {\it size function} $\Phi(r_{1/2},z)$ measures the number of galaxies per unit comoving volume and per unit
half-light radius $r_{1/2}$. In order to chart the evolution of the number density of disk galaxies with a given size as a function of redshift up to $z\sim1$, we split our sample of disk galaxies into four different redshift bins of width $\Delta z=0.2$, centred on the redshifts 0.3, 0.5, 0.7 and 0.9. 

There are two different considerations in constructing the size function which should be clearly distinguished. The first is the variation of sampling volume for different individual objects that are included in the sample. This can be fully taken into account using the $V_{max}$-formalism (\cite{schmidt68}; \cite{felten76}) in which each galaxy is weighted with the reciprocal of the comoving volume it could occupy while still satisfying the selection criteria of the sample. For bright objects this volume will be larger, thus leading to a smaller weight and correcting for their relative rarity with respect to fainter members of the galaxy population:
\begin{equation}
\Phi(r_{1/2},z)\times\Delta r_{r_{1/2}}=\Sigma_{\gamma}~\frac{1}{V_{max,\gamma}}~.
\end{equation}   
The summation is carried out over all sources $\gamma$ with a photometric redshift within the redhift bin in consideration.  Using that
the comoving distance $D(z)$ in a $\Lambda$CDM cosmological model with $\Omega_{\Lambda}=0.75$ and $\Omega_{m}=0.25$ is given by
\begin{equation}
D(z)=\left(\frac{c}{H_0}\right)~\int^z_0\frac{dz}{\sqrt{\Omega_{\Lambda}+\Omega_{m}~(1+z)^3}}~,
\end{equation}
the volume $V_{max}$ is calculated for each source according to:
\begin{equation}
V_{max}=\left(\frac{c}{H_0}\right)^3~d\Omega~\left[\int^{z_{max}}_{z_{min}}\frac{dz}{\sqrt{\Omega_{\Lambda}+\Omega_{m}~(1+z)^3}}\right]^3~.
\end{equation}
Here $d\Omega$ is the effective solid angle covered by the survey during Cycle 12 which is 2734 arcmin$^2$.

The lower and upper bounds of integration, $z_{min}$ and $z_{max}$, define the redshift range within which any particular galaxy in the sample would have entered the sample. It is constrained in three ways: (a) the minimum and maximum redshifts of the relevant redshift bin; (b) the minimum and maximum redshifts at which the sample magnitude limits are satisfied, and (c) the minimum and maximum redshifts at which the size limits in the sample are fulfilled. The size limit of relevance is that the object must have sufficient pixels to be classified by $ZEST$ (FWHM $> 0.15$ arcsec) 
and can  be calculated from the condition that the galaxy have the same physical size $r_{1/2}\propto D_A\cdot R_{1/2}[px]$ at all redshifts, regardless of its observed scale ($\frac{D_A(z_{max,~size})}{D_A(z_{obs})}=\frac{R_{1/2}}{R_{1/2,~min}}$).
In practice the size constraint is the least significant because of the small variation in angular diameter with redshift at high redshifts and because of the high angular resolution of the HST images. Of more importance are the limits in redshift imposed by the magnitude cuts. In the case of a bounding magnitude of $I_{lim}$ -- in the present analysis either $I=16$ or $I=22.5$ -- the requirement of identical absolute magnitude leads to the condition: $M_B=I_{lim}-5~log\left(\frac{D_L(z_{lim})}{10 pc}\right)+K_{F814W,B}(z)$, 
which must be solved for the limiting redshift $z_{lim}$\footnote{Here $K_{F814W,~B}$ denotes the K-correction as defined through the equation $I_{AB}=M_B+5~log\left(D_L(z)/10~pc\right)+K_{I,B}$ (c.f. \cite{hogg02}).}.

There may also be an observational surface brightness selection. It is, however, unimportant in practice when compared with the surface
brightness constraint induced by a selection in absolute magnitude. The difficulty is that the size function is only partially sampled,
because of the magnitude limits of the sample. Thus the size function constructed above represents the integral of the bivariate
size-luminosity function above some limiting luminosity. We adopt the approach of \cite{lil98} to deal with this problem. We (a) limit the
sample in equation (2) to objects lying within a specified range of luminosity, and allow this luminosity range to change with redshift
according to the expected luminosity evolution of individual galaxies. The integration over luminosity of the bivariate function is thus
explicit, albeit model dependent. Secondly, we consider, (b), at each redshift, the limiting size above which one can be fairly confident that the sample is more or less ``complete'', i.e. that there are rather few objects of this size, or greater, lying below the magnitude limit because of their low surface brightness. Of course, luminous low surface brightness galaxies are known to exist and the sample can never be absolutely complete, but we may assume that the size function above this limiting size will be a good approximation to reality and it will, in any case, represent a lower limit to the actual number density. It is important to appreciate that this apparent surface brightness selection arises from the inevitable selection in absolute magnitude (at a given size) rather than any direct observational surface brightness selection (c.f. \cite{lil98} for a detailed discussion of this matter). 

Figure 12 shows the distribution of $M_B$ and $r_{1/2}$ in the sample as a function of redshift. As discussed in the previous paragraph, we limit the set of galaxies to those above a redshift-dependent magnitude limit of $M_B = (-19.6 - z)$. It reflects an assumed luminosity evolution by about one magnitude due to a passively aging stellar population since $z\sim1$. The horizontal dotted lines in Figure 12 illustrate the effect of the cut as it would be applied at the central redshift of the different bins. Their loci change from an absolute magnitude of $M_B=-20.5$ at redshift 0.9 to $M_B=-19.6$ at redshift zero. In Figure 12 the inclined dotted lines are lines of constant surface brightness, shifted according to the assumed surface brightness evolution. The figure illustrates that our sample is fairly complete above $r_{1/2} \sim$ 5 kpc, i.e. the bulk of galaxies with sizes above this threshold lies above the horizontal line at all redshifts. We therefore limit our analysis to galaxies with half-light radii exceeding this size.

\subsection{Errors}

Sizes, magnitudes and photometric redshifts are all subject to uncertainty. In order to assess the resulting inaccuracy of the size function $\Phi(r_{1/2},z)$, we varied these quantities according to a Gaussian distribution centred on the measured values and having an appropriate standard deviation. In the case of the photometric redshifts, for instance, the standard deviation rises with increasing redshift as $\sigma_z\approx 0.03\cdot(1+z)$. For the magnitude, we used the standard deviation $\sigma_I$ given in the ACS-COSMOS catalog. 
Errors for the sizes are tabulated for each galaxy by GIM2D in the form of upper and lower 99\% confidence levels, corresponding to 3$\sigma_{R_{1/2}}$ for the half-light radii. 

From the results of 101 test runs with errors introduced on all three of the above quantities, we calculated the resulting scatter for each bin of the size function. The size of the errors given on the grounds of these tests corresponds to the interquartile range of the 101 values at every measurement point on the curve $\Phi(r_{1/2},z)$. The interquartile range spans 50\% of a data set, and eliminates the influence of outliers by effectively removing the highest and lowest quarters of the data range. The distance from the median of the 101 error realisations to the first and third quartile give the lower and upper error bar respectively. In the plots of the size function (c.f. Figures 13 and 14) these asymmetric errors have been reported in each individual data point and the area between them shaded in grey.

The Poisson noise in each bin of the size function (owing to the statistical manner in which each bin is populated) remains, however, the dominant source of random error in the size function. The Poisson error is given by:
\begin{equation}
\sigma_P = \sqrt{\sum_i \frac{1}{V_{max,i}^2}}
\end{equation}
with the summation carried out over all objects $i$ in a given bin. Poisson errors are reported symmetrically around the data points in each bin
of the size function.

Systematic uncertainties in the GIM2D measurements of the half-light radii (discussed in Appendix A1) could affect the
slope of the size functions presented in Figures 13 and 14 for the COSMOS galaxies, and in Figures 16 and 17 for our comparison sample of SDSS galaxies
(section 4.4 below). It should be noted that any systematic errors arising from the fits should be identical at $z = 0$ and $z = 0.7$ and thus the {\it SDSS-normalized}
densities of Figures 18 and 19 - on which we base our discussion and conclusions - are free of such systematics.

\subsection{Comparison with the local SDSS galaxy population}

To quantitatively assess the redshift evolution of the number density of disk galaxies of a given size (and as a function of $B/D$-ratio),
it is important to realize that a large overdensity in the COSMOS field at $z < 0.4$ affects the number counts in the lowest COSMOS redshift
bin. Therefore, to carry out our comparison and determine how the number of disk galaxies with scalelengths larger than about 5 kpc changes
with redshift from $z\sim1$ to $z\sim0$, we repeated our analysis on a set of 1876 galaxies extracted from the SDSS Data Release 4. To fully
include all observational effects, the images of the SDSS galaxies were degraded so as to appear as they would in the COSMOS images at $z =
0.7$. The generation of these  simulated images will be described in detail elsewhere (\cite{kampczyk06}). To summarize, the SDSS galaxies
were selected from the redshift range $z\in[0.015, 0.025]$. Their $g$-band images were then transformed to how such galaxies would appear in
the F814W COSMOS ACS images if they were to lie at $z = 0.7$, at which redshift the passbands are well matched. Thus, this ``redshifting" of
the SDSS galaxies needed only to take into account the different pixel scales, and point spread function, and the cosmological surface
brightness dimming. Two sets of images were produced, one without surface brightness evolution and one with an assumed brightening to high
redshift of $\Delta \mu = z$ magnitudes arcsec$^{-2}$. (We will henceforth refer to these two samples as ``unbrightened" and ``brightened",
respectively). No size evolution was considered. The simulated galaxies at $z = 0.7$ were then added into the COSMOS ACS images to reproduce
also the same issues of image crowding and noise. These $z=0.7$-``simulated" SDSS galaxies were then analyzed following the identical procedure adopted for the real COSMOS objects. Stamps sized $10''\times10''$ or more were extracted for all galaxies.  As before, the original images were cleaned if necessary and then symmetrised prior to performing the GIM2D fits. After applying exactly the same selection criteria with respect to magnitude and $\chi_{GIM2D}^2$, 457 and 1294 objects remained in the unbrightened and brightened samples, respectively. The distribution of absolute B-band magnitudes, half-light radii and S\'ersic indices among the two sets of artificially redshifted SDSS galaxies is compared to that of COSMOS galaxies in the range $z\in[0.6,0.8[$ in Figure 15. 
 
Absolute B-band magnitudes for the SDSS galaxies were derived by determining the best fit spectral energy distribution template with ZEBRA, and then calculating the $K$-correction for that spectral type, and the corresponding distance modulus, at redshift $z\sim0.7$. The SDSS sample of \cite{kampczyk06} was extracted as a volume limited sample from the local Universe; its equivalent volume was obtained by computing the luminosity function for the 1876 SDSS galaxies, and setting it equal to the global SDSS luminosity function. We used the so-derived equivalent volume to compute the size function for the SDSS sample.

\section{Results and discussion}

Figures 13 and 14 show the size functions for the total disk galaxy population and for disk galaxies of different $B/D$-ratios, derived as a function of redshift above an evolving absolute magnitude limit $M_{B} = (-19.6 - z)$ (c.f. Section 4.2). As discussed before, these size functions should be most complete for sizes larger than roughly 5 kpc. Above this threshold we have thus fit them with straight lines of $log(\Phi)$ vs. $r_{1/2}$, i.e. with an exponential size function.  The fits of the aforementioned functional form were calculated for two sets of photometric ZEBRA redshifts; one obtained without (solid line) and the other with (dashed line) corrections to the photometric catalog (with no significant change to our conclusions). The reported size function is based on the uncorrected photometry and the parameters of the corresponding solid black lines are given in Table 4. The size function of disk galaxies as a whole appears remarkably constant with redshift across the range $0.2 < z < 1.0$. We detect, however, a weak trend for the slope of the size function to steepen with redshift, a result which applies to the global disk population and also separately to all sub-classes of bulge-to-disk ratio.  

These findings are strengthened by comparing directly the COSMOS size function in the range $z\in[0.6,0.8[$ with that derived from the local SDSS sample  (``redshifted" to $z\sim0.7$). In Figures 16 and 17, the SDSS-based size functions are plotted together with the COSMOS size functions in the bin $z\in[0.6,0.8[$. The size function of the brightened SDSS sample is reported with black points and error bars (and labeled as `SDSS$^*$'); in grey we show the unbrightened one (`SDSS'). The difference is due to the brightening of the galaxies which, in combination with the luminosity cut at $M_{B} \sim -20.3$ at z = 0.7 (c.f. Figure 12) leads to a greater number of smaller objects in the brightened set of SDSS galaxies. Given the evidence for luminosity evolution in all classes of galaxies, we would expect the brightened sample to provide the best comparison and in Figure 15 the latter indeed matches the distributions of $M_B$, $n$ and $r_{1/2}$ of COSMOS galaxies best. The line parameters of a linear regression on the SDSS size function in the same range of sizes as in the COSMOS data set are listed in Table 5.
Figures 16 and 17 show a striking similarity in the shape of the size function over half the Hubble time. Furthermore, the steepening of the size function in COSMOS relative to locally is confirmed in that there is a small deficit of the largest disks ($r_{1/2} > 7$ kpc) in the COSMOS sample compared to the artificially redshifted local sample.

To further investigate this issue we show in the two following figures the density of disk galaxies in COSMOS normalized to the local size function
of (brightened) SDSS sources for two size bins, for all disks (Figure 18) and differentiated by bulge to disk ratio (Figure 19). In Table 6
we list the relative densities $\widetilde{\rho}=\Phi_{COSMOS} /  \Phi_{SDSS^*}$ for disk galaxies as a function of redshift and
$B/D$-category. Here $\Phi_{COSMOS}$ denotes the comoving number density per unit size, as presented in Figures 13
and 14 for the COSMOS sample in the studied redshift bins. $\Phi_{SDSS^*}$ stands for the (artificially redshifted and brightened) SDSS density in the
$0.6-0.8$ redshift bin (as presented in Figure 16 for the entire disk galaxy sample, and in Figure 17 for the sample split in the four B/D ratio
bins).  In the table, $\widetilde{\rho}_<$ stands for the relative density of disks with intermediate sizes of $r_{1/2}\in$ [5 kpc, 7 kpc[ and
$\widetilde{\rho}_>$ for that of large disks  with $r_{1/2}\in$ [7 kpc, 17 kpc[. At $z=0$, the normalized density point $\Phi_{SDSS^*}/\Phi_{SDSS^*}=1$
has been plotted as well, so as to show the error bar on the SDSS $z=0$ density estimate.  Within the uncertainties (and excluding the lowest redshift
bin from COSMOS because of the large cosmic variance), there is no evidence for a variation with redshift in the size function of disks in the range
$r_{1/2} \in$ [5 kpc, 7 kpc[. Disk galaxies of these intermediate sizes have neither grown nor been destroyed from $z=1$ to $z=0$ or, alternatively,
the growth and destruction of such systems exactly balanced over this period. The larger disks, however, show a weak decline by a factor of nearly two
over the same time span, even though the sample should be more complete for these larger disks as discussed before.

Figures 14, 17 and especially Figure 19 and the values of the slopes $\alpha$ given in Table 4, consistently show the steepening of the
slope of the size function to be the more pronounced for galaxies with higher $B/D$-ratios. This suggests that the deficit of large disks at
earlier epochs is primarily due to a deficit of {\it bulged} disk galaxies. In principle disks with the highest bulge-to-disk ratio could be
classified as early-types by ZEST and thus excluded from our sample, yet the direct comparison with the SDSS sample - which was classified
in precisely the same fashion - argues against this interpretation. Moreover, the stated result remains valid even if  type 2.0 galaxies are
omitted from the discussion.  Morphological $K$-corrections  could play a  role in  the observed
disappearance of about 60\% of the large disks with bulges, due to color differences between bulges and disks; we are studying this issue  in more detail.

The constancy - or even slight increase - of the number density of {\it bulgeless} disks out to $z\sim 1$ is at variance with models based on
hierarchical formation or infall scenarios (e.g., \cite{mao98}; \cite{bouwenssilk02}). On the other hand, a constant or slightly increasing number
density of large bulgeless disks at $z\sim1$, in combination with a decrease of similarly sized {\it bulged} disks to this redshift, supports a scenario
where secular evolution processes form or at least grow the sizes of  bulges at late stages in the history of the universe. Independent evidence for
bulge-building secular evolution within disk galaxies comes from numerical experiments (e.g., \cite{norman96}; Debattista et al.\ (2004, 2005,
2006)), observations of nearby ``young" bulges with disk-like stellar density profiles (see \cite{kormendy1} and \cite{carollo2004} for reviews and
references; \cite{carollo1999}, Carollo et al.\ (1998, 2001, 2002), \cite{carollo2006}), and also from the lack of evolution in the fraction of
large-scale bars in disk galaxies since $z=1$ (Sheth et al.\ (2003); Jogee et al (2004)). Our work provides strong evidence that, by $z\sim1$, the
disks have not only grown to their current sizes, but also that they have become  massive and stable enough to allow for large-scale internal
dynamical instabilities to take place and to  grow bulges inside disks, thereby shaping the Hubble sequence that we observe today.

\section{Conclusions}

We have used measurements of the half-light radii obtained with analytic fits to the surface brightness distribution of roughly 12,000  disk galaxies in COSMOS to chart the evolution of the number density of intermediate-sized and large disks since redshift $z\sim1$. 

Our analysis has shown a general constancy of the size function for disk galaxies with half-light radii larger than about 5 kpc. This constancy, initially proposed by (\cite{lil98}) based on much smaller statistics, and here clearly demonstrated from a much larger sample of disk galaxies of any bulge-to-disk ($B/D$) ratio, suggests that the massive disk galaxy population was largely in place with its current properties when the universe was only half as old as today. 

Among the observed general constancy of the size function there is evidence for some subtler and possibly quite important trends: $(i)$ The
number density of disk galaxies of all $B/D$-ratios in the intermediate size range $r_{1/2}\in$ [5 kpc, 7 kpc[ remains constant to a good
approximation up to $z\sim1$. $(ii)$ The number density of the largest disks with half-light radii $r_{1/2}\in$ [7 kpc, 17 kpc[ at $z\sim1$
appears however to be only 60\% of today's value in the disk galaxy population as a whole. $(iii)$ When this population of large-sized disk
galaxies is split by the values of their $B/D$-ratio we find that the deficit at high redshift stems from disk galaxies with a bulge
component.  It is conceivable that color differences between bulges and disks contribute to this result
and we will study their effect in more detail.  Bearing this in mind, however, it is nevertheless likely that we might be  witnessing a secular transformation of bulgeless disks into disks with a bulge component.

\clearpage

\appendix 

\section{Reliability of the GIM2D fits}

We have conducted several tests to assess the reliability of the GIM2D fits. It must be emphasized that even objects with a high value of
$\chi_{GIM2D}^2$ represent successful fits to the underlying symmetric component of the surface brightness profile in most cases. The
$\chi^2$ value supplied by GIM2D is the sum over all pixels within the fitted area of the squared difference between model and observed data.
Assuming that the pixel noise is purely due to Poisson photon statistics, this sum is then normalised to the variance of the pixel noise.
Figure 1 clearly shows that high $\chi_{GIM2D}^2$ values mostly reflect the presence of spiral arms and other features of galactic structure
that cannot be captured by the single-S\'ersic models, which continue however to provide a good description of the underlying galaxy light.

\subsection{Simulations}

To quantitatively test the robustness of the GIM2D fits, we performed a set of simulations of an equal number of  pure exponential disks ($n=1$) and
pure de Vaucouleur galaxies ($n=4$) with a range of half-light radii and magnitudes (and ellipticities) covering the observed parameter space.  

It should be emphasized that the vast majority of disk galaxies in our sample is well described by a $n=1$ S\'ersic profile (c.f. Figure 9) and that none of the (disk-selected) galaxies in our
sample are well represented by a smooth pure de Vaucoleur profile which would be appropriate for a disk-less spheroid. By construction, the ZEST-selected disk galaxies of our study posess a visible disk
component, even if the single-S\'ersic fits that we have performed return an $n=4$ profile. Nonetheless, we include the extreme pure de Vaucouleur
cases in our simulation sample, to derive  conservative estimates for the systematic errors in our GIM2D measurements.

The simulated galaxies were generated with the IRAF task \textit{mkobjects},
which automatically adds a component of Poisson noise to the flux of the source. Source-independent background noise was estimated from five
regions of pure sky in different ACS-tiles of the COSMOS survey. An average standard deviation from these regions was used to scale a Gaussian
white noise distribution, in order to match the average background characteristics of the COSMOS images.  The Sextractor measurements and the GIM2D
fits for the simulated galaxies were carried out in an analogous way to the measurements and fits of the COSMOS galaxies (c.f. Section 2). 

Figures 20 and 21 summarize the main results derived from the simulated galaxies. Figure 20 displays the distribution of  normalized difference
between the input and the recovered S\'ersic indices $n$, as a function of  galaxy magnitude and separately for the $n=1$ and the $n=4$ galaxies. Figure
21 shows the comparison between the input half-light radii and those derived from the GIM2D fits; in particular, the figure shows the distribution
of the normalized difference between GIM2D and input half-light radii - separately for the $n=1$ and the $n=4$ galaxies as in the previous figure - as a function of magnitude
(upper half) and of galaxy size (lower half). 

The GIM2D fits tend to underestimate the galaxy sizes. There is a dependence on   galaxy magnitude and size, and on the shape of the profile:  For
example, the radii are underestimated by 3\% and 10\% for $n=1$, $I_{AB}>21$ galaxies with half-light radii smaller or equal and larger than 10 pixels,
respectively. The similar comparison for the $n=4$ galaxies yields systematic uncertainties of order 20\% and 40\%, respectively.   

These systematic  effects could affect the slopes of  the size functions presented in Figures 13 and 14 for the COSMOS galaxies, and in Figures 16 and
17 for the comparison sample of SDSS galaxies. 
We are, however, confident that these systematic errors on the GIM2D half-light radii should not affect our conclusions. First, an influence of such
systematics on the measured size function would actually cause an {\it increase} of the number density of large bulgeless systems at $z\sim 1$.
Second, and most important, the approach of using a sample of SDSS galaxies redshifted to $z$=0.7 to anchor the COSMOS size function to the local
values ensures that these potential systematic errors on the size measurements for any disk galaxy type - including the most centrally concentrated,
largest and/or faintest disk galaxies in our sample - should be identical at $z = 0$ and $z = 0.7$.

\subsection{Susceptibility to PSF Variations}

Although we did not expect the outcome of single S\'ersic fits to depend strongly on small variations of the PSF, we carried out tests with
varying PSFs on a few COSMOS galaxies. The outcome of such a test for one of these objects is illustrated in Figure 22. The six panels show
the distribution of parameters put out by GIM2D if the galaxy is fitted with 250 different PSFs extracted in a (spatially) regular fashion
from the total of 2500 PSFs generated by \cite{rhod06} for each ACS tile to study weak lensing maps. The results of the PSF tests confirm that
the GIM2D fits are very robust to possible variations of the PSF as occur due to spatial variations across the ACS-tiles, for changes in the
S\'ersic index and half-light radius are well below the 10\% level.

\acknowledgments
We thank the anonymous referee for his valuable comments and suggested improvements to the original manuscript.\\
The HST COSMOS Treasury program was supported through NASA grant HST-GO-09822. We wish to thank Tony Roman, Denise Taylor, and David
Soderblom for their assistance in planning and scheduling of the extensive COSMOS observations. We gratefully acknowledge the contributions
of the entire COSMOS collaboration consisting of more than 70 scientists.  More information on the COSMOS survey is available at\\
http://www.astro.caltech.edu/\raisebox{-0.9ex}{\~{ }}cosmos. We acknowledge the services provided by the NASA IPAC/IRSA
staff in providing online archive and server capabilities for the COSMOS
datasets.\\
P. Kampczyk, C. Scarlata and M.~T. Sargent acknowledge support from the Swiss National Science Foundation.

{\it Facilities:} \facility{HST (ACS)}.

\clearpage

\begin{deluxetable}{ccccccccc}
\tabletypesize{\scriptsize}
\tablecaption{Physical properties of the galaxies shown in Figures 1 to 7$^a$. \label{tbl-1}}
\tablewidth{0pt}
\tablehead{
\colhead{redshift} & \colhead{$M_B$} & \colhead{$I_{AB}$} & \colhead{size $[kpc]$} & \colhead{$n$} & \colhead{type} & \colhead{$e$} & \colhead{$\chi^2_{GIM2D}$} }
\startdata
0.271 & -21.558 & 18.303 & 11.17 & 1.19 & 2.0 & 0.36 & 7.651\\
0.097 & -18.008 & 19.740 & 2.06 & 0.55 & 2.0 & 0.77 & 5.378\\
0.340 & -21.604 & 18.765 & 9.56 & 0.78 & 2.0 & 0.30 & 5.364\\
0.511 & -21.465 & 19.882 & 3.43 & 4.06 & 1 & 0.09 & 1.148\\
0.572 & -20.762 & 20.933 & 4.21 & 5.48 & 1 & 0.35 & 1.059\\
1.085 & -22.453 & 21.881 & 4.76 & 3.45 & 1 & 0.23 & 1.044\\
0.751 & -22.890 & 19.808 & 9.89 & 0.90 & 2.3 & 0.05 & 1.640\\
0.212 & -18.525 & 20.914 & 3.28 & 0.90 & 2.3 & 0.46 & 1.242\\
0.850 & -21.100 & 22.016 & 5.65 & 0.68 & 2.3 & 0.23 & 1.031\\
0.215 & -19.500 & 19.894 & 4.99 & 1.19 & 2.2 & 0.33 & 1.167\\
0.075 & -15.416 & 20.964 & 1.92 & 1.22 & 2.2 & 0.63 & 1.014\\
0.854 & -20.951 & 22.179 & 3.86 & 1.47 & 2.2 & 0.44 & 1.016\\
0.391 & -20.440 & 20.023 & 8.55 & 2.00 & 2.1 & 0.86 & 1.176\\
0.742 & -21.678 & 20.958 & 7.59 & 2.48 & 2.1 & 0.43 & 1.181\\
1.080 & -22.371 & 21.913 & 5.40 & 3.04 & 2.1 & 0.54 & 1.065\\
0.755 & -22.543 & 20.179 & 7.49 & 1.40 & 2.0 & 0.33 & 2.164\\
0.729 & -22.636 & 19.911 & 13.45 & 1.81 & 2.0 & 0.42 & 2.505\\
0.343 & -19.272 & 20.830 & 1.71 & 4.21 & 2.0 & 0.72 & 1.207\\
0.300 & -20.363 & 19.912 & 5.17 & 1.67 & 3 & 0.24 & 1.584\\
0.649 & -21.299 & 20.906 & 6.18 & 1.36 & 3 & 0.48 & 1.124\\
0.682 & -20.542 & 21.826 & 4.77 & 0.62 & 3 & 0.44 & 1.040\\
\enddata
\tablenotetext{a}{~The galaxies are listed in the order in which they appear in Figures 1 to 7, i.e. the first galaxy listed is the uppermost one in Figure 1, the third one the lowermost object in Figure 1, the fourth entry the uppermost galaxy in Figure 2, and so forth.
}

\end{deluxetable}

\clearpage

\begin{deluxetable}{lccccccc}
\tabletypesize{\scriptsize}
\tablecaption{Number of galaxies  as a function of redshift when a 2$\sigma$-cut in $\chi_{phz}^2$ \label{tbl-2} is applied}
\tablewidth{0pt}
\tablehead{
\colhead{redshift} & \colhead{total sample} &  \colhead{type 2.3} & \colhead{type 2.2} & \colhead{type 2.1} & \colhead{type2.0} & \colhead{type 1} & \colhead{type 3}}
\startdata
$0.2 \leq z <0.4$ & 3967 & 813 & 1733 & 571 & 276 & 425& 132 \\
$0.4 \leq z <0.6$ & 2634& 714 & 969 & 346 & 188 & 256 & 153 \\
$0.6 \leq z <0.8$ & 3313 & 813 & 873 & 506 & 357 & 540& 219 \\
$0.8 \leq z <1$ & 2508 & 761 & 493 & 347 & 234 & 416& 249 \\
$1 \leq z <1.2$ & 527 & 167 & 114 & 77 & 32 & 54& 82 \\
\enddata
\end{deluxetable}

\begin{deluxetable}{lccccccc}
\tabletypesize{\scriptsize}
\tablecaption{Number of galaxies  as a function of redshift without a 2$\sigma$-cut in $\chi_{phz}^2$ \label{tbl-3}}
\tablewidth{0pt}
\tablehead{
\colhead{redshift} & \colhead{total sample} &  \colhead{type 2.3} & \colhead{type 2.2} & \colhead{type 2.1} & \colhead{type2.0} & \colhead{type 1} & \colhead{type 3}}
\startdata
$0.2 \leq z <0.4$ & 4180 & 841 & 1791 & 601 & 309 & 465& 149 \\
$0.4 \leq z <0.6$ & 2859 & 758 & 1045 & 381 & 214 & 279& 171 \\
$0.6 \leq z <0.8$ & 3616 & 871 & 953 & 559 & 387 & 583& 248 \\
$0.8 \leq z <1$ & 2840 & 800 & 552 & 415 & 287 & 507& 270 \\
$1 \leq z <1.2$ & 661 & 188 & 134 & 96 & 52 & 93& 96 \\
\enddata
\end{deluxetable}

\clearpage

\begin{deluxetable}{ccccccccc}
\tabletypesize{\scriptsize}
\tablecaption{Line parameters of the COSMOS size function for radii $r_{1/2}>5$ kpc$^a$.\label{tbl-4}}
\tablewidth{0pt}
\tablehead{
\colhead{redshift} & \colhead{type} & \colhead{$\alpha$} & \colhead{$\beta$} & \colhead{$\chi^2$} }
\startdata
0.3 & type 2 & -0.200 $\pm$ 0.021 & -2.278 $\pm$ 0.167 &  0.912\\
0.5 & type 2 & -0.215 $\pm$ 0.024 & -2.494 $\pm$ 0.180 & 0.834\\
0.7 & type 2 & -0.240 $\pm$ 0.017 & -2.225 $\pm$ 0.125 & 5.363\\
0.9 & type 2 & -0.294 $\pm$ 0.031 & -1.891 $\pm$ 0.215 & 0.311\\
\hline
0.3 & type 2.3 & -0.253 $\pm$ 0.070 & -2.602 $\pm$ 0.501 & 0.204\\
0.5 & type 2.3 & -0.221 $\pm$ 0.071 & -2.910 $\pm$ 0.485 & 0.279\\
0.7 & type 2.3 & -0.283 $\pm$ 0.034 & -2.321 $\pm$ 0.237 & 2.940\\
0.9 & type 2.3 & -0.291 $\pm$ 0.049 & -2.130 $\pm$ 0.331 & 0.800\\
\hline
0.3 & type 2.2 & -0.234 $\pm$ 0.039 & -2.372 $\pm$ 0.293 & 0.457\\
0.5 & type 2.2 & -0.236 $\pm$ 0.039 & -2.735 $\pm$ 0.291 & 1.641\\
0.7 & type 2.2 & -0.204 $\pm$ 0.022 & -2.913 $\pm$ 0.171 & 6.963\\
0.9 & type 2.2 & -0.260 $\pm$ 0.035 & -2.696 $\pm$ 0.254 & 4.164\\
\hline
0.3 & type 2.1 & -0.149 $\pm$ 0.061 & -3.468 $\pm$ 0.500 & 0.088\\
0.5 & type 2.1 & -0.171 $\pm$ 0.050 & -3.710 $\pm$ 0.446 & 0.060\\
0.7 & type 2.1 & -0.198 $\pm$ 0.043 & -3.446 $\pm$ 0.330 & 1.562\\
0.9 & type 2.1 & -0.214 $\pm$ 0.046 & -3.528 $\pm$ 0.377 & 0.155\\
\hline
0.3 & type 2.0 & -0.125 $\pm$ 0.029 & -3.585 $\pm$ 0.303 & 0.981\\
0.5 & type 2.0 & -0.129 $\pm$ 0.031 & -4.031 $\pm$ 0.326 & 2.274\\
0.7 & type 2.0 & -0.218 $\pm$ 0.046 & -3.304 $\pm$ 0.354 & 0.082\\
0.9 & type 2.0 & -0.242 $\pm$ 0.068 & -3.458 $\pm$ 0.533 & 0.136\\
\enddata
\tablenotetext{a}{~Line parameters and $\chi^2$ of the linear regressions of the form $log(\Phi)=\alpha \cdot r_{1/2} + \beta$ applied to the size functions of Figures 13 and 14 at sizes $r_{1/2}>$5 kpc (solid black lines in the figures). The values of slope $\alpha$ and y-axis intercept $\beta$ are given for the morphological class type 2 as a whole, and for the different $B/D$-ratio sub-classes. See Figures 13 and 14 for the graphical representation of the size function derived from the COSMOS galaxies.
}
\end{deluxetable}

\clearpage

\begin{deluxetable}{ccccccccc}
\tabletypesize{\scriptsize}
\tablecaption{Line parameters of the size function derived from artificially redshifted SDSS galaxies for radii $r_{1/2}>5$ kpc$^a$.\label{tbl-5}}
\tablewidth{0pt}
\tablehead{
\colhead{sample} & \colhead{type} & \colhead{$\alpha$} & \colhead{$\beta$} & \colhead{$\chi^2$} }
\startdata
SDSS & type 2 & -0.159 $\pm$ 0.027 & -3.051 $\pm$ 0.234 &  2.589\\
SDSS & type 2.3 & -0.172 $\pm$ 0.056 & -3.449 $\pm$ 0.494 & 0.429\\
SDSS & type 2.2 & -0.110 $\pm$ 0.034 & -3.877 $\pm$ 0.341 & 1.942\\
SDSS & type 2.1 & -0.128 $\pm$ 0.080 & -4.013 $\pm$ 0.675 & 1.351\\
SDSS & type 2.0 & -0.100 $\pm$ 0.304 & -4.158 $\pm$ 2.046 & 0.002\\
\hline
SDSS$^*$ & type 2 & -0.197 $\pm$ 0.026 & -2.470 $\pm$ 0.201 &  0.939\\
SDSS$^*$ & type 2.3 & -0.176 $\pm$ 0.040 & -3.060 $\pm$ 0.326 & 2.434\\
SDSS$^*$ & type 2.2 & -0.217 $\pm$ 0.055 & -2.727 $\pm$ 0.404 & 0.402\\
SDSS$^*$ & type 2.1 & -0.107 $\pm$ 0.050 & -4.031 $\pm$ 0.503 & 0.244\\
SDSS$^*$ & type 2.0 & -0.142 $\pm$ 0.048 & -3.743 $\pm$ 0.431 & 1.683\\
\enddata
\tablenotetext{a}{~Line parameters and $\chi^2$ of the linear regressions of the form $log(\Phi)=\alpha \cdot r_{1/2} + \beta$ applied to the size functions of the brightened (denoted as `SDDS$^*$' in the table) and unbrightened (denoted `SDDS') artificially redshifted SDSS galaxies at sizes $r_{1/2}>$5 kpc. The values of slope $\alpha$ and y-axis intercept $\beta$ are given for the morphological class type 2 as a whole, and for the different $B/D$-ratio sub-classes. See Figures 16 and 17 for the graphical representation of the size function derived from the SDSS galaxies.
}
\end{deluxetable}

\clearpage

\begin{deluxetable}{ccccccccc}
\tabletypesize{\scriptsize}
\tablecaption{Relative densities $\widetilde{\rho}=\Phi_{COSMOS}/\Phi_{SDSS^*}$ as a function of redshift$^a$.\label{tbl-6}}
\tablewidth{0pt}
\tablehead{
\colhead{redshift} & \colhead{type} & \colhead{$\widetilde{\rho}_<$} & \colhead{$\widetilde{\rho}_>$}}
\startdata
0.3 & type 2 & 1.386  $\pm$ 0.327 & 1.664  $\pm$ 0.513\\
0.5 & type 2 & 0.710  $\pm$ 0.163 & 0.686  $\pm$ 0.214\\
0.7 & type 2 & 0.960  $\pm$ 0.196 & 0.681  $\pm$ 0.197\\
0.9 & type 2 & 0.920  $\pm$ 0.195 & 0.616  $\pm$ 0.185\\
\hline
0.3 & type 2.3 & 0.904  $\pm$ 0.398 & 0.798  $\pm$ 0.502\\
0.5 & type 2.3 & 0.710  $\pm$ 0.274 & 0.610  $\pm$ 0.336\\
0.7 & type 2.3 & 1.169  $\pm$ 0.393 & 0.706  $\pm$ 0.349\\
0.9 & type 2.3 & 1.550  $\pm$ 0.528 & 1.142  $\pm$ 0.551\\
\hline
0.3 & type 2.2 & 1.639  $\pm$ 0.585 & 2.379  $\pm$ 1.241\\
0.5 & type 2.2 & 0.737  $\pm$ 0.262 & 0.819  $\pm$ 0.445\\
0.7 & type 2.2 & 0.832  $\pm$ 0.270 & 0.811  $\pm$ 0.411\\
0.9 & type 2.2 & 0.632  $\pm$ 0.215 & 0.372  $\pm$ 0.205\\
\hline
0.3 & type 2.1 & 1.954  $\pm$ 1.392 & 1.311  $\pm$ 0.934\\
0.5 & type 2.1 & 0.792  $\pm$ 0.573 & 0.582  $\pm$ 0.415\\
0.7 & type 2.1 & 1.089  $\pm$ 0.706 & 0.446  $\pm$ 0.305\\
0.9 & type 2.1 &  0.707  $\pm$ 0.473 & 0.311  $\pm$ 0.217\\
\hline
0.3 & type 2.0 & 1.473  $\pm$ 0.903 & 2.651  $\pm$ 1.973\\
0.5 & type 2.0 & 0.578  $\pm$ 0.364 & 0.724  $\pm$ 0.584\\
0.7 & type 2.0 & 0.725  $\pm$ 0.403 & 0.681  $\pm$ 0.513\\
0.9 & type 2.0 & 0.338  $\pm$ 0.209 & 0.345  $\pm$ 0.279\\
\enddata
\tablenotetext{a}{~Relative densities $\widetilde{\rho}=\Phi_{COSMOS} /  \Phi_{SDSS^*}$ for disk galaxies as a function of redshift and
$B/D$-category. Here $\Phi_{COSMOS}$ and $\Phi_{SDSS^*}$ denominate the comoving number density per unit size as reported in the plots of the
size function in Figures 13 and 14 for the COSMOS sample in the redshift range $z\in[0.6,0.8[$, and in Figures 15 and 17 for the artificially redshifted and brightened sample of the SDSS galaxies. $\widetilde{\rho}_<$ stands for the relative density of disks with intermediate sizes of $r_{1/2}\in[5~kpc, 7~kpc[$ and $\widetilde{\rho}_>$ for that of large disks with $r_{1/2}\in[7~kpc, 17~kpc[$.}
\end{deluxetable}

\clearpage

\begin{figure}[ht]
\epsscale{1.0}
\plotone{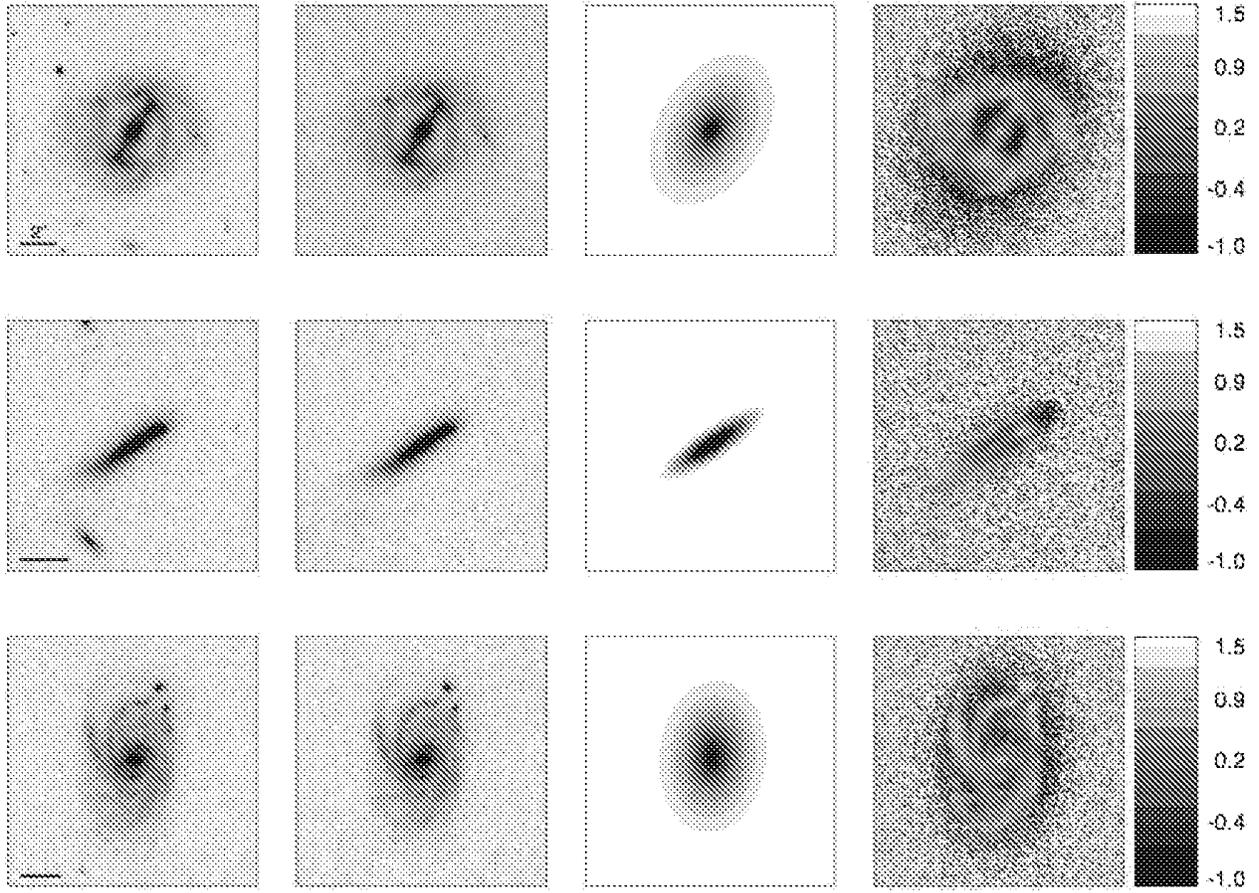}
\caption{\label{bigchi} Selection of objects with high values of $\chi_{GIM2D}^2$ but satisfactory fits to the smooth
component of the surface brightness profile of the galaxy. From left to right, the figure shows the original stamp
image, the cleaned image, the GIM2D model and the residual image (c.f. section 3). For each object, the  bar in the original galaxy image indicates a
scale of  2 arcseconds. An intensity scale for the normalized residual images is reported on their right-hand side. Model parameters are listed in Table 1.}
\end{figure}

\clearpage

\begin{figure}[ht]
\epsscale{1.0} 
\plotone{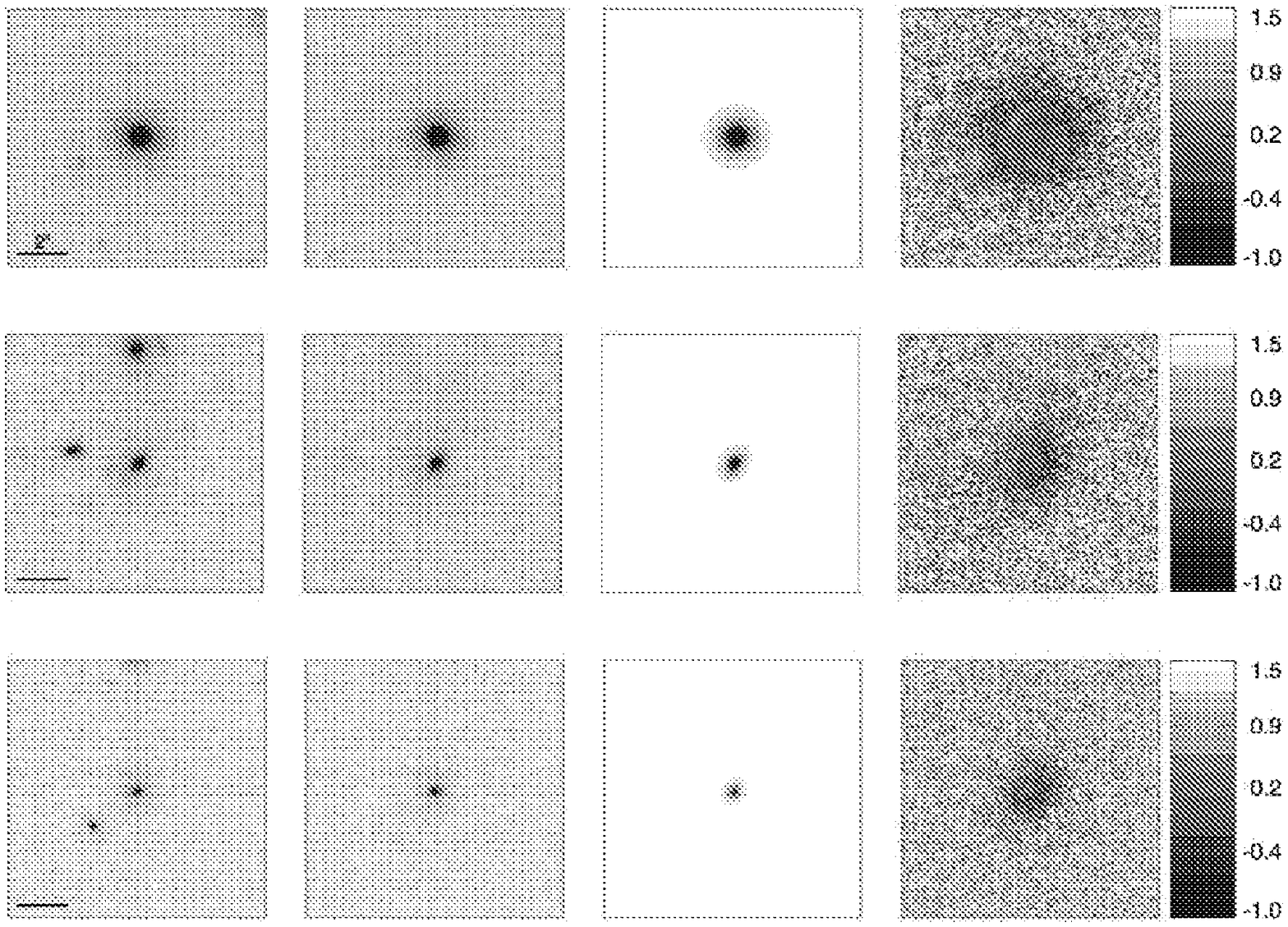}
\caption{\label{type1a} Original images, cleaned images, GIM2D model and residual image for three galaxies of ZEST
morphological type 1. Color and size scales are as in Figure 1; the values of the physical parameters are listed in Table 1.}
\end{figure}

\clearpage

\begin{figure}[ht]
\epsscale{1.0}
\plotone{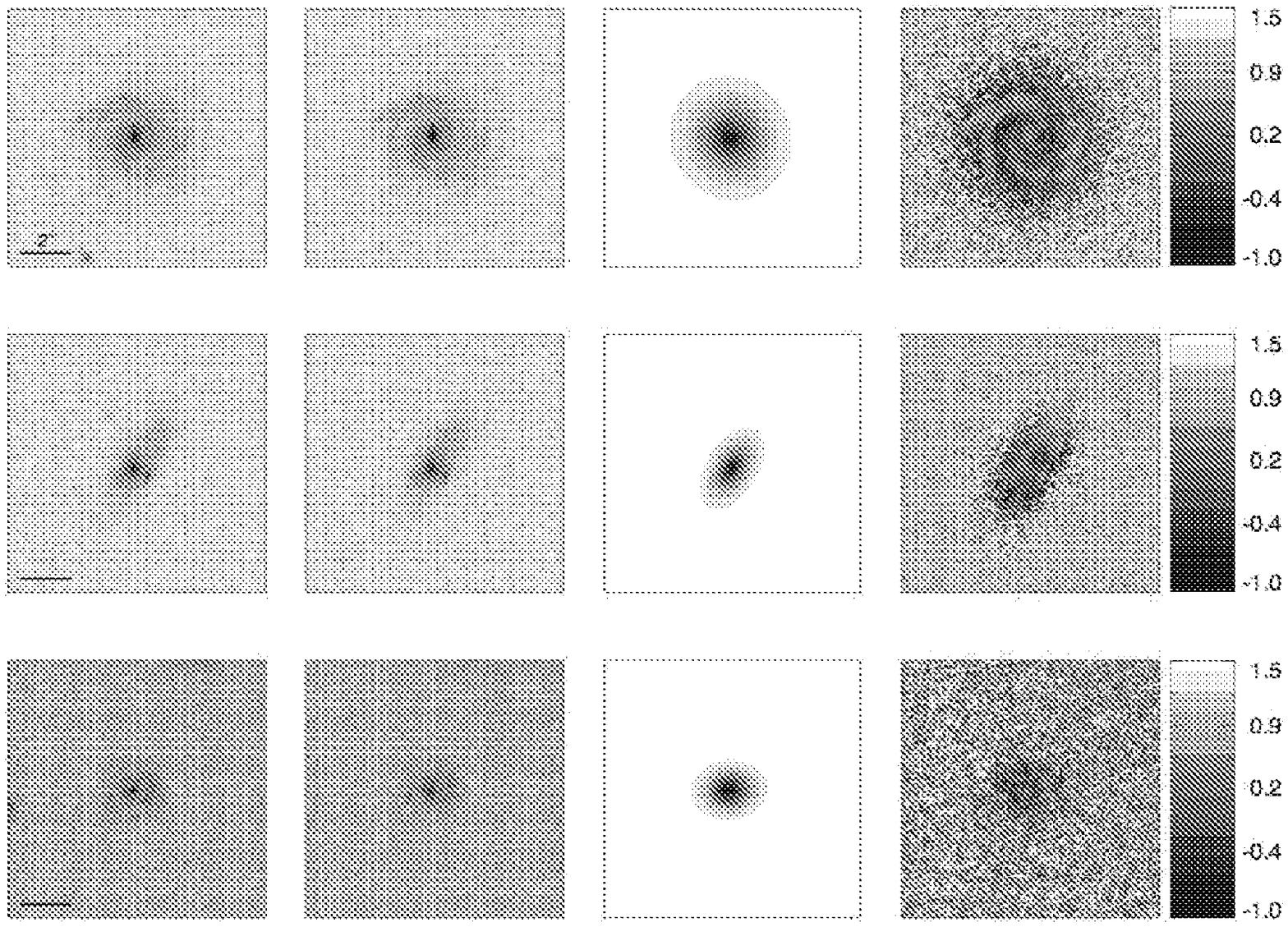}
\caption{\label{type2pt0a} Original images, cleaned images, GIM2D model and residual image for three galaxies of ZEST
morphological type 2.3. Color and size scales are as in Figure 1; the values of the physical parameters are listed in Table 1.}
\end{figure}

\clearpage

\begin{figure}[ht]
\epsscale{1.0}
\plotone{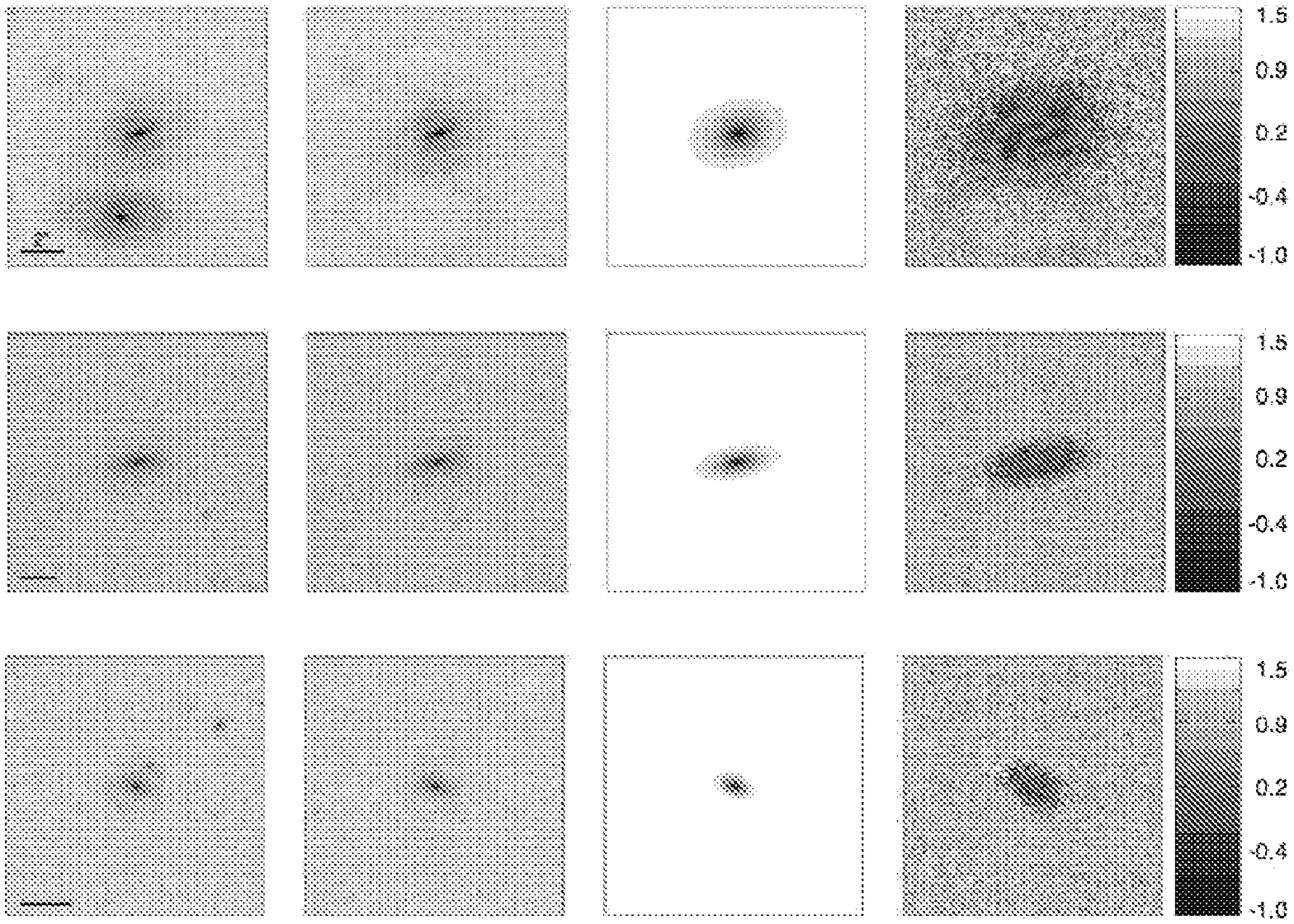}
\caption{\label{type2pt1a} Original images, cleaned images, GIM2D model and residual image for three galaxies of ZEST morphological type
2.2. Color and size scales are as in Figure 1; the values of the physical parameters are listed in Table 1.}
\end{figure}

\clearpage

\begin{figure}[ht]
\epsscale{1.0} 
\plotone{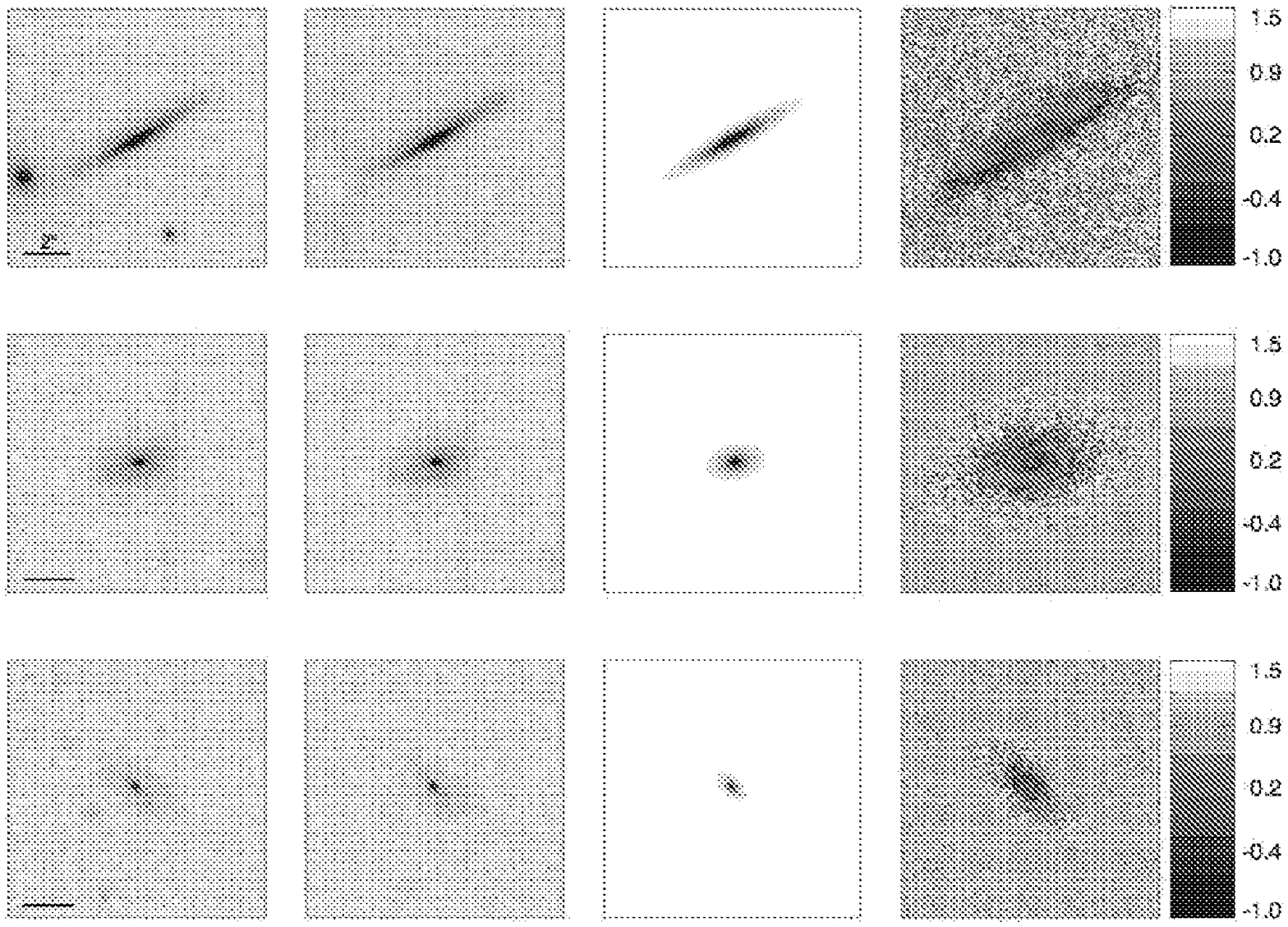}
\caption{\label{type2pt2a} Original images, cleaned images, GIM2D model and residual image for three galaxies of ZEST morphological type
2.1. Color and size scales are as in Figure 1; the values of the physical parameters are listed in Table 1.}
\end{figure}

\clearpage

\begin{figure}[ht]
\epsscale{1.0}
\plotone{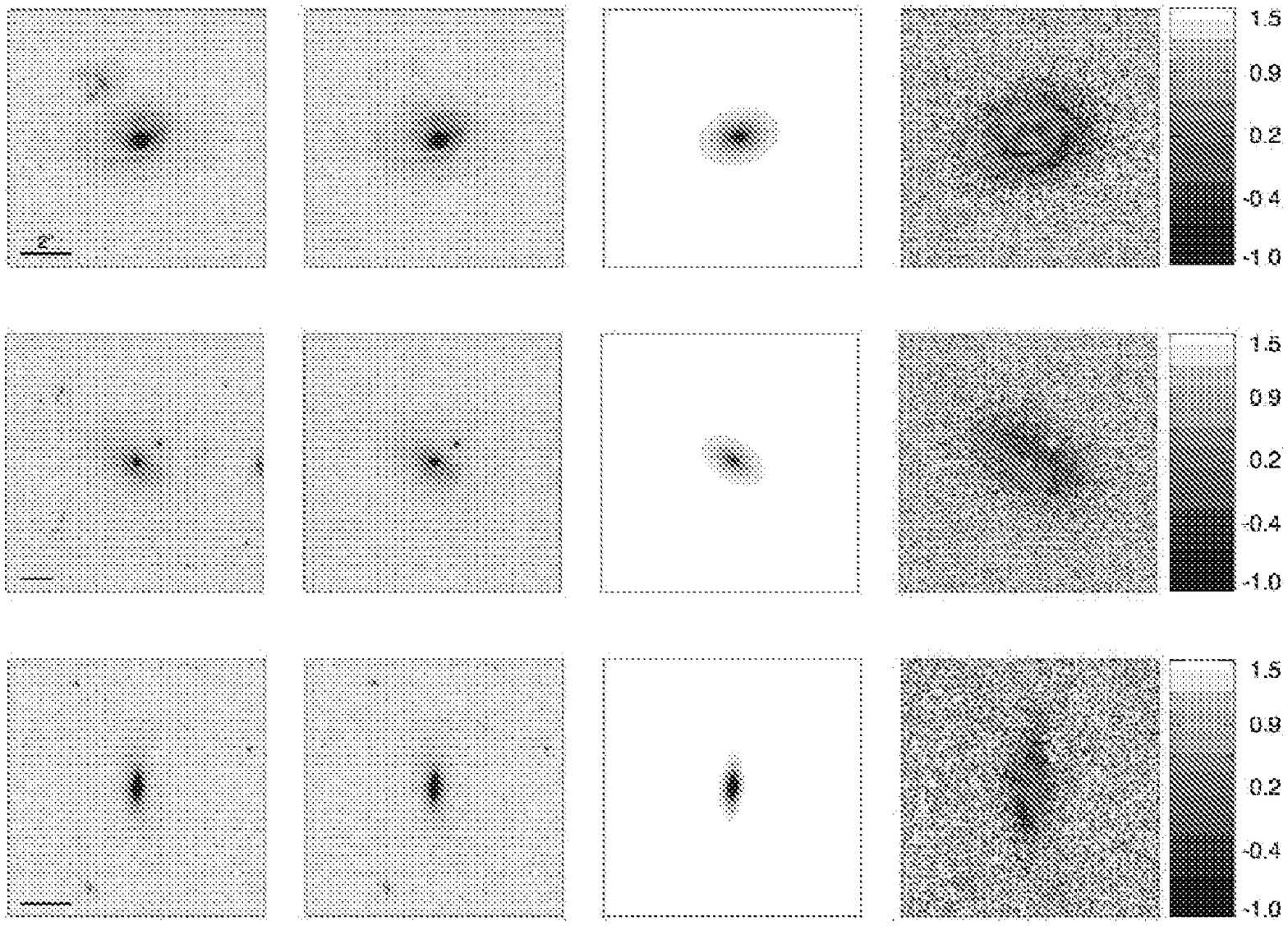}
\caption{\label{type2pt3a} Original images, cleaned images, GIM2D model and residual image for three galaxies of ZEST morphological type
2.0. Color and size scales are as in Figure 1; the values of the physical parameters are listed in Table 1.}
\end{figure}

\clearpage

\begin{figure}[ht]
\epsscale{1.0}
\plotone{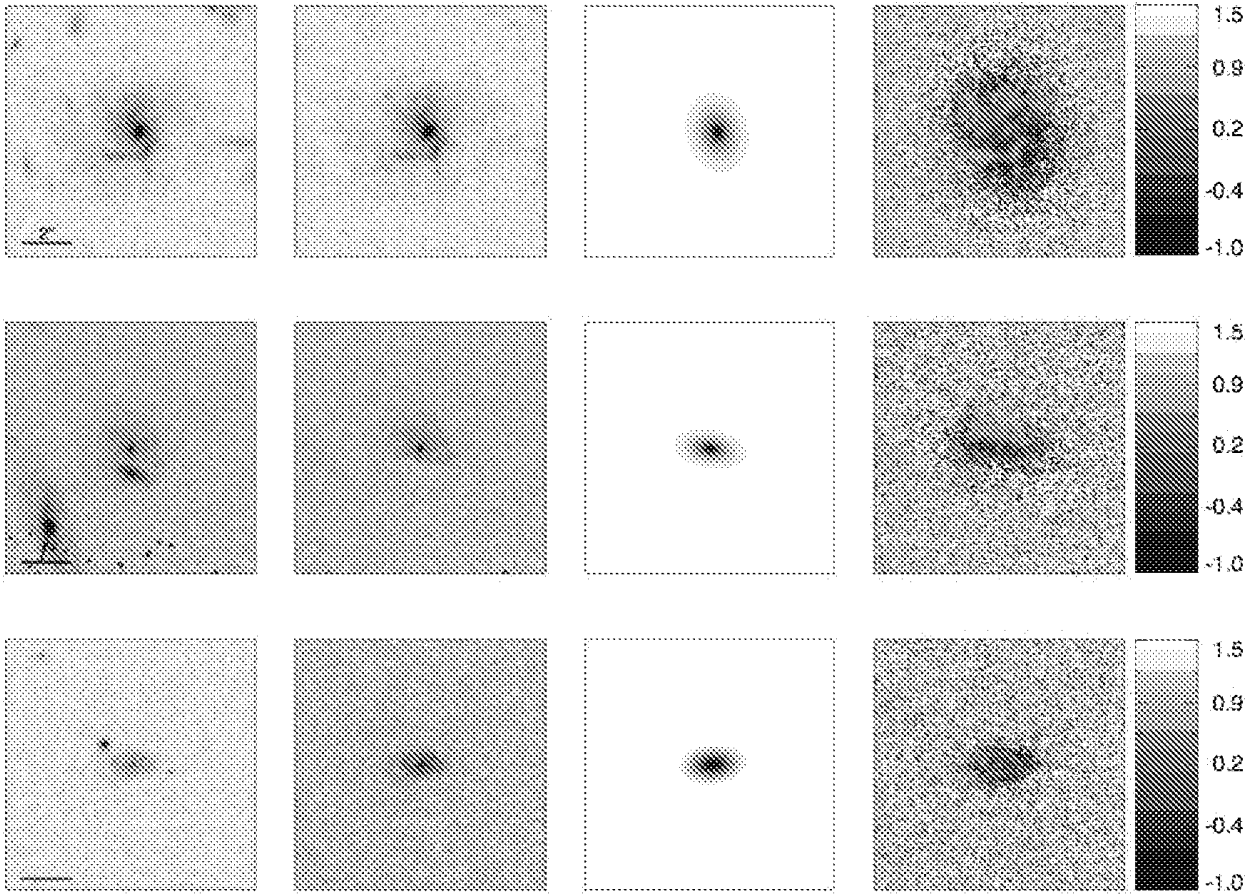}
\caption{\label{type3a} Original images, cleaned images, GIM2D model and residual image for three galaxies of ZEST morphological type 3.
Color and size scales are as in Figure 1; the values of the physical parameters are listed in Table 1. Due to the irregularity of the
distribution of light in this class of objects, these GIM2D fits are inevitably the most unreliable.} 
\end{figure}

\clearpage

\begin{figure}[ht]
\epsscale{1.0}
\plotone{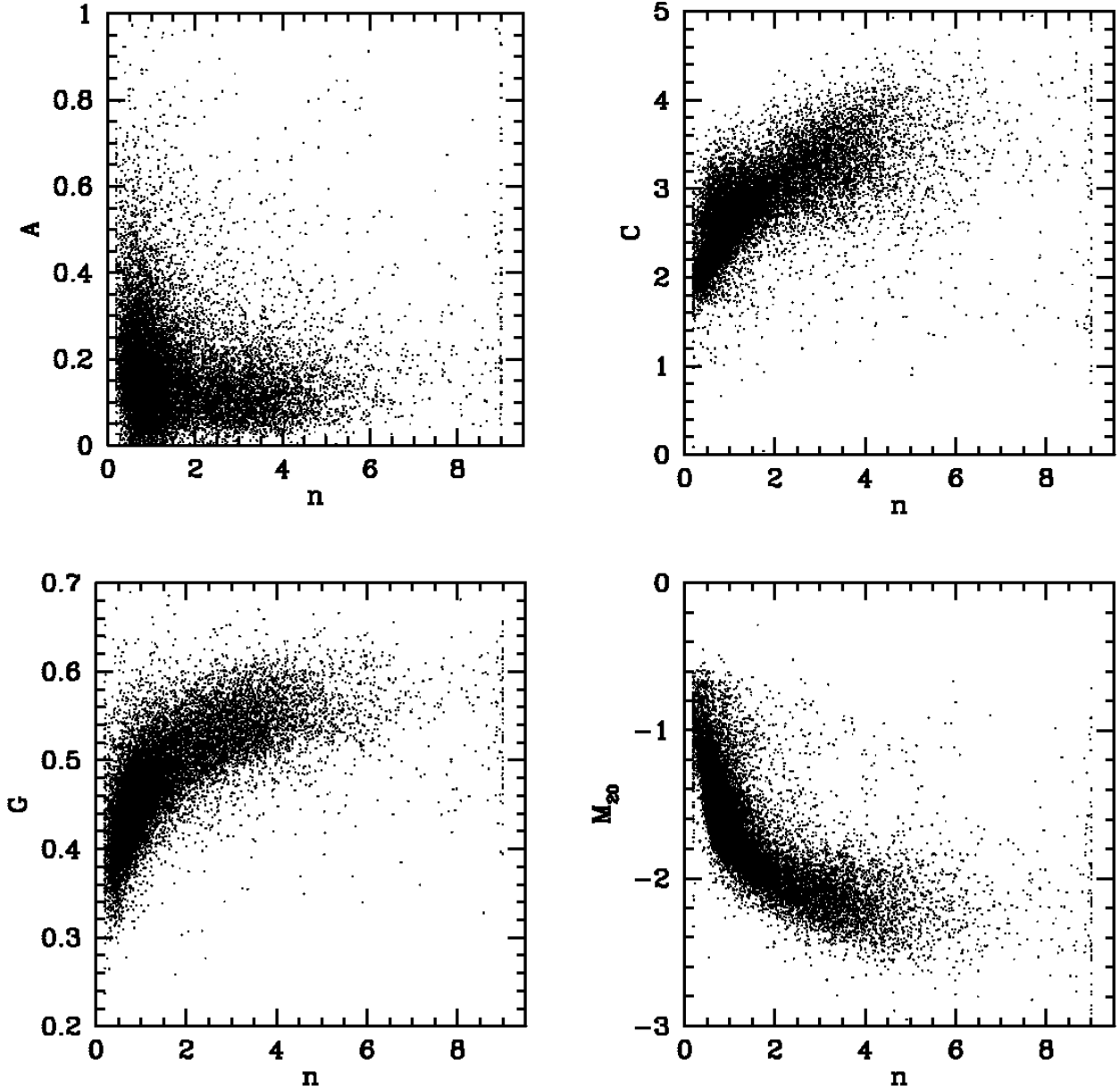}
\caption{\label{comparisonnonparam} The comparison of the non-parametric diagnostics $A$ (asymmetry), $C$ (concentration), $M_{20}$ (second order moment of the brightest 20\% of the pixels) and $G$ (Gini coefficient) with the S\'ersic index $n$. The regular trends and tight relations in these plots support the reliability of the GIM2D fits. The clustering of points at $n=9$ reflects our restriction of the S\'ersic index to the interval $n\in[0.2,9]$.}
\end{figure}

\clearpage

\begin{figure}[ht]
\epsscale{1.0}
\plotone{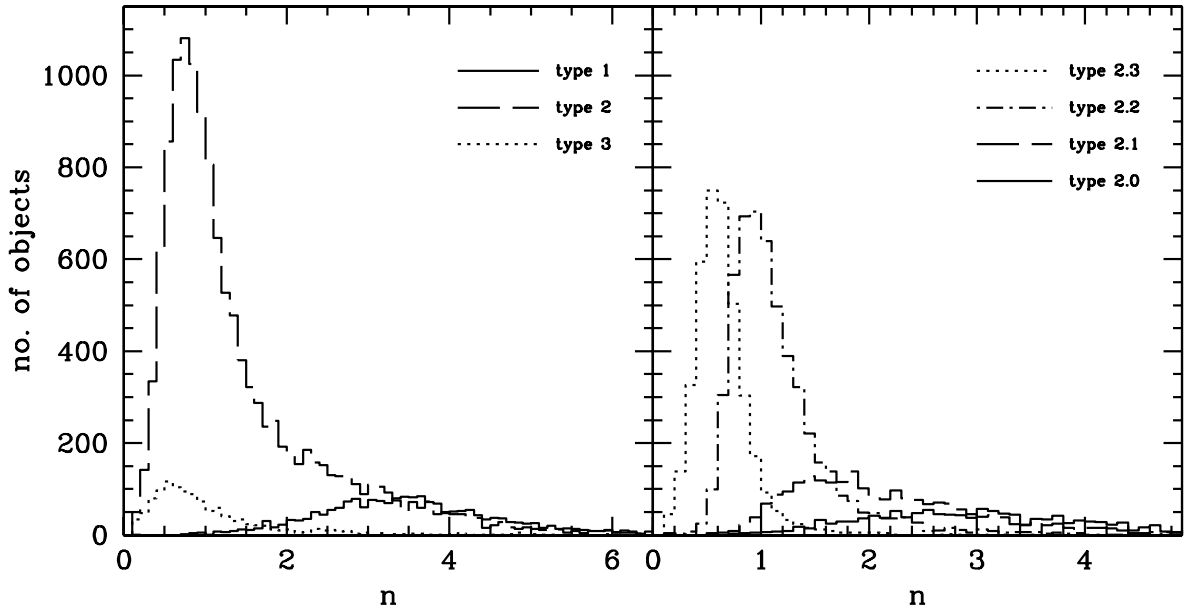}
\caption{\label{ndistr}Left half: Distribution of S\'ersic indices for galaxies classified by ZEST as type 1 (early-types), type 2 (disk galaxies) and type 3 (irregular and peculiar galaxies). Right half: Distribution of S\'ersic indices for sources ranging from purely exponential disks (type 2.3) to bulge-dominated disks (type 2.0).}
\end{figure}

\clearpage

\begin{figure}[ht]
\epsscale{1.0}
\plotone{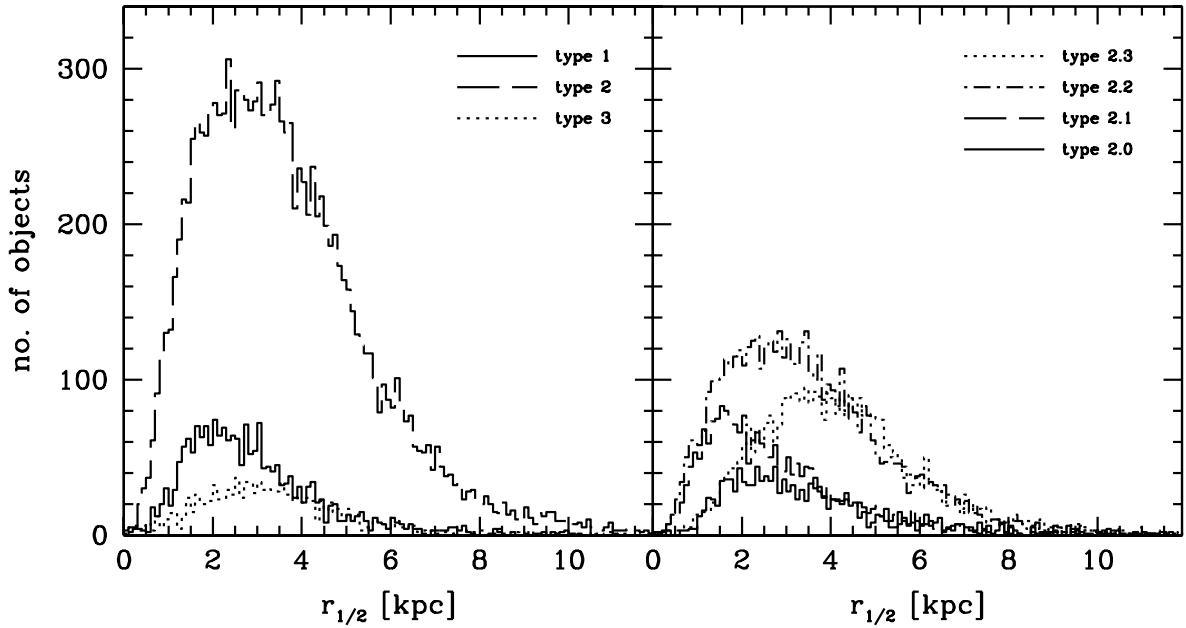}
\caption{\label{Rdistr}Left half: Distribution of half-light radii (measured in kpc) for galaxies classified by ZEST as type 1, type 2 and type 3. Right half: Distribution of half-light radii for objects ranging between pure exponential disks (type 2.3) and strongly bulge-dominated objects (type 2.0).}
\end{figure}

\clearpage

\begin{figure}[ht]
\epsscale{1.0}
\plotone{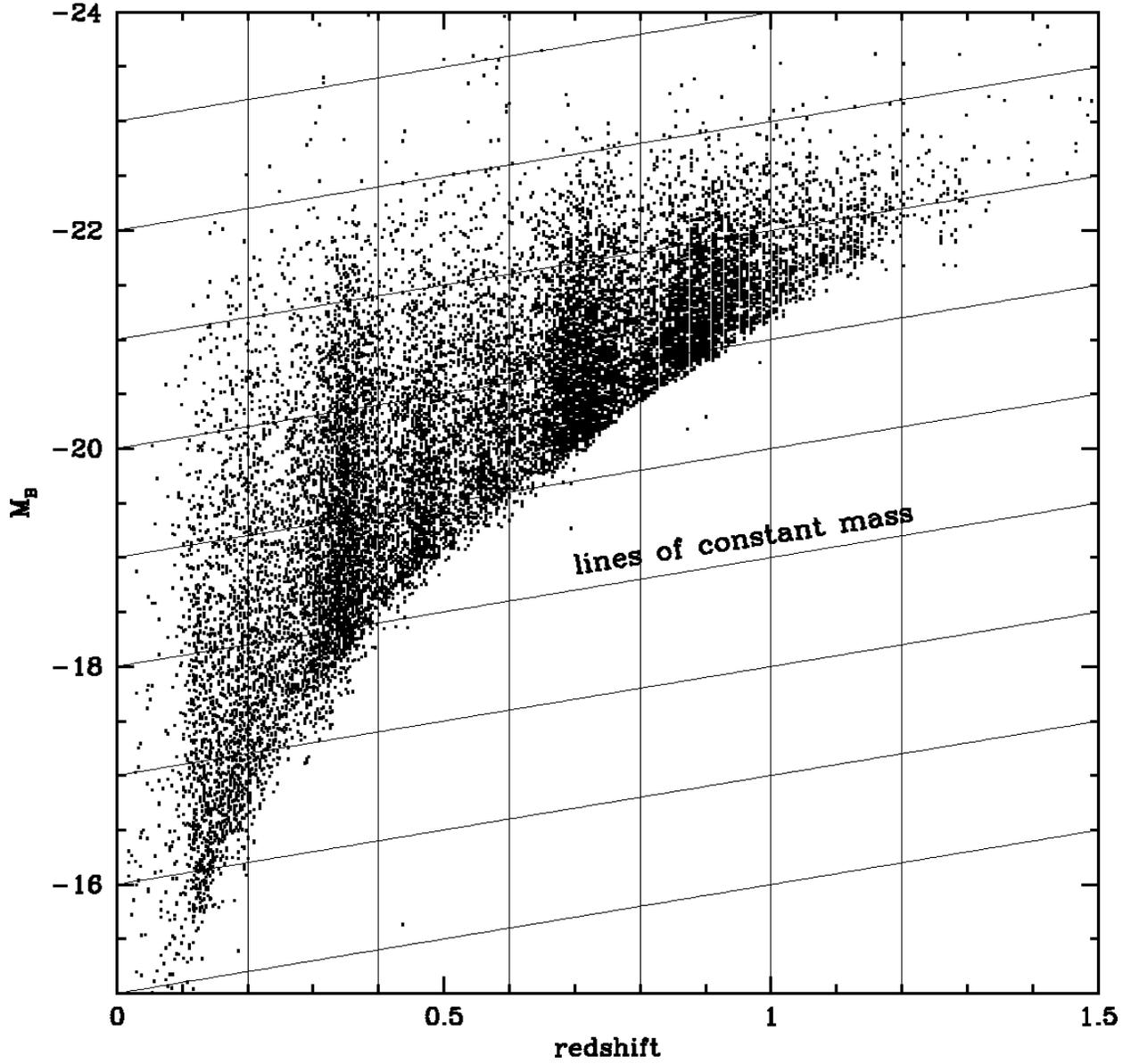}
\caption{\label{zmb} The distribution of all Cycle 12 COSMOS galaxies with $I\in[16,22.5]$ and absolute $B$-band magnitude $M_B$ at redshifts $z<1.5$.}
\end{figure}

\clearpage

\begin{figure}[ht]
\epsscale{1.0}
\plotone{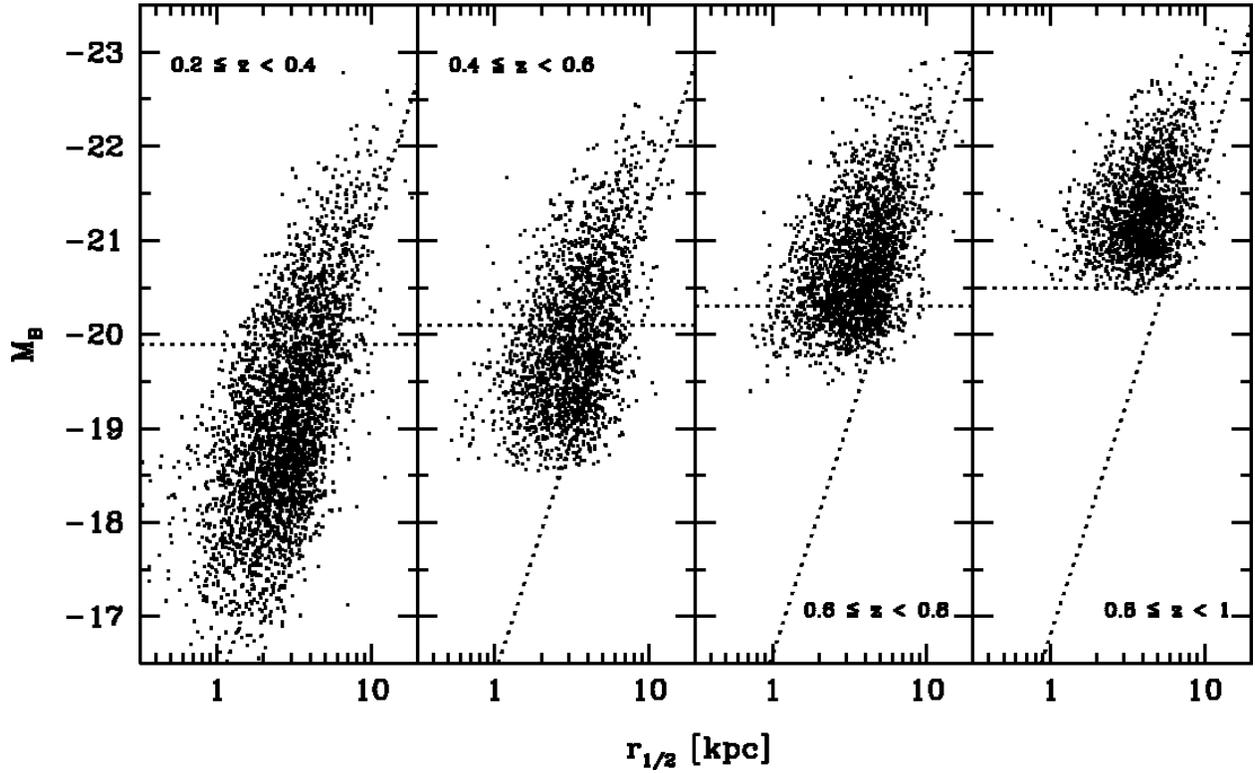}
\caption{Distribution of absolute $B$-band magnitude $M_B$ vs. physical size for type 2 disk galaxies as a function of redshift. Horizontal lines show the average magnitude cuts applied in the derivation of the size function within the different redshift bins (c.f. Section 4.2) and inclined dotted lines are lines of constant surface brightness.}
\end{figure}

\clearpage

\begin{figure}[ht]
\epsscale{0.7} 
\plotone{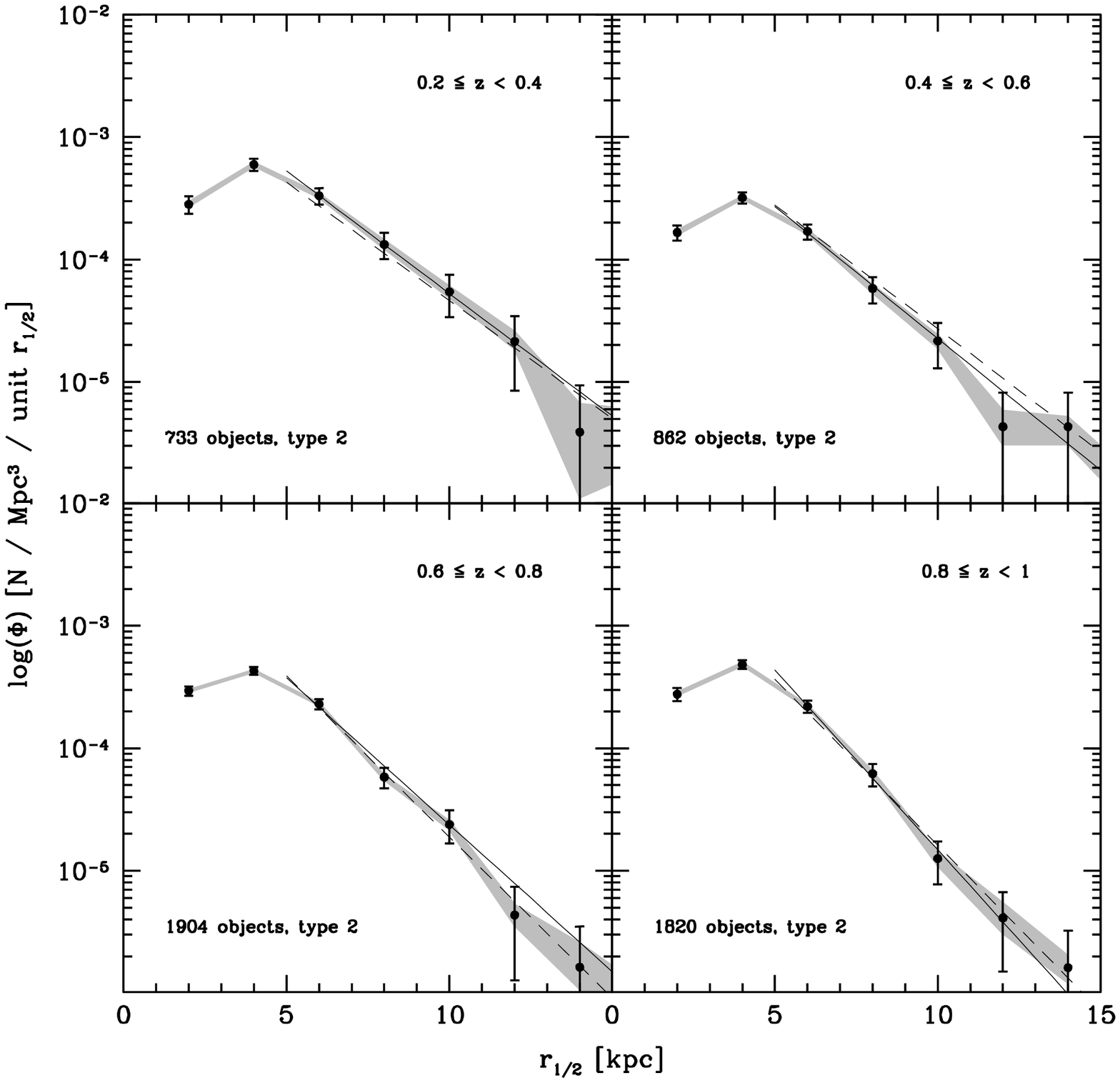}
\caption{Evolution of the size function between redshift 0.2 and 1 for the entire sample of disk galaxies. Four different bins of redshift, centred at
$z\sim$ 0.3, 0.5, 0.7 and 0.9, are displayed. The vertical error bars reproduce  the Poisson errors. Areas shaded in grey indicate the errorbar on the
size function induced by random errors in redshift, size and observed magnitude (c.f. Section 4.3). The solid and dashed lines in black show fits to the relation $log(\Phi)=\alpha \cdot r_{1/2} + \beta$ above
5 kpc, calculated for two sets of ZEBRA photometric redshifts obtained without and with corrections to the photometric catalogs, respectively (c.f.
Section 4.1; Table 4 lists the best fit parameters of the solid lines).\newline
Apart from a modest steepening of the slope with increasing redshift, the size function of disk galaxies is roughly constant in the range $0.2 < z < 1.0$.}
\end{figure}

\clearpage

\begin{figure}[ht]
\epsscale{0.8} 
\plotone{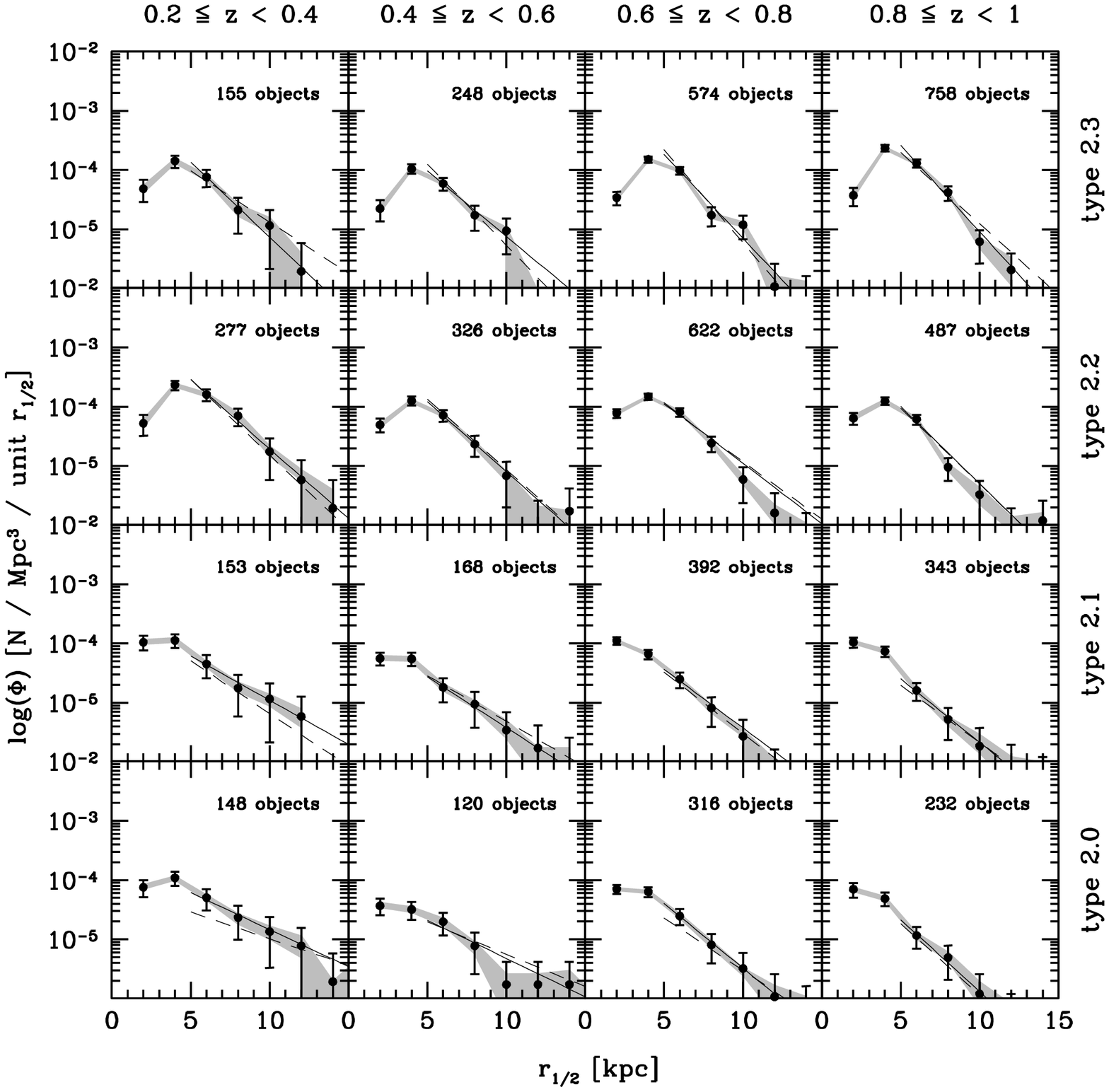}
\caption{The evolution of the size function shown individually for all sub-classes of the disk galaxy population in COSMOS. The sub-classes are displayed by increasing $B/D$-ratio from top to bottom. The errors and the solid and dashed black lines are as in Figure 13.}
\end{figure}

\clearpage

\begin{figure}[ht]
\epsscale{1.0}
\plotone{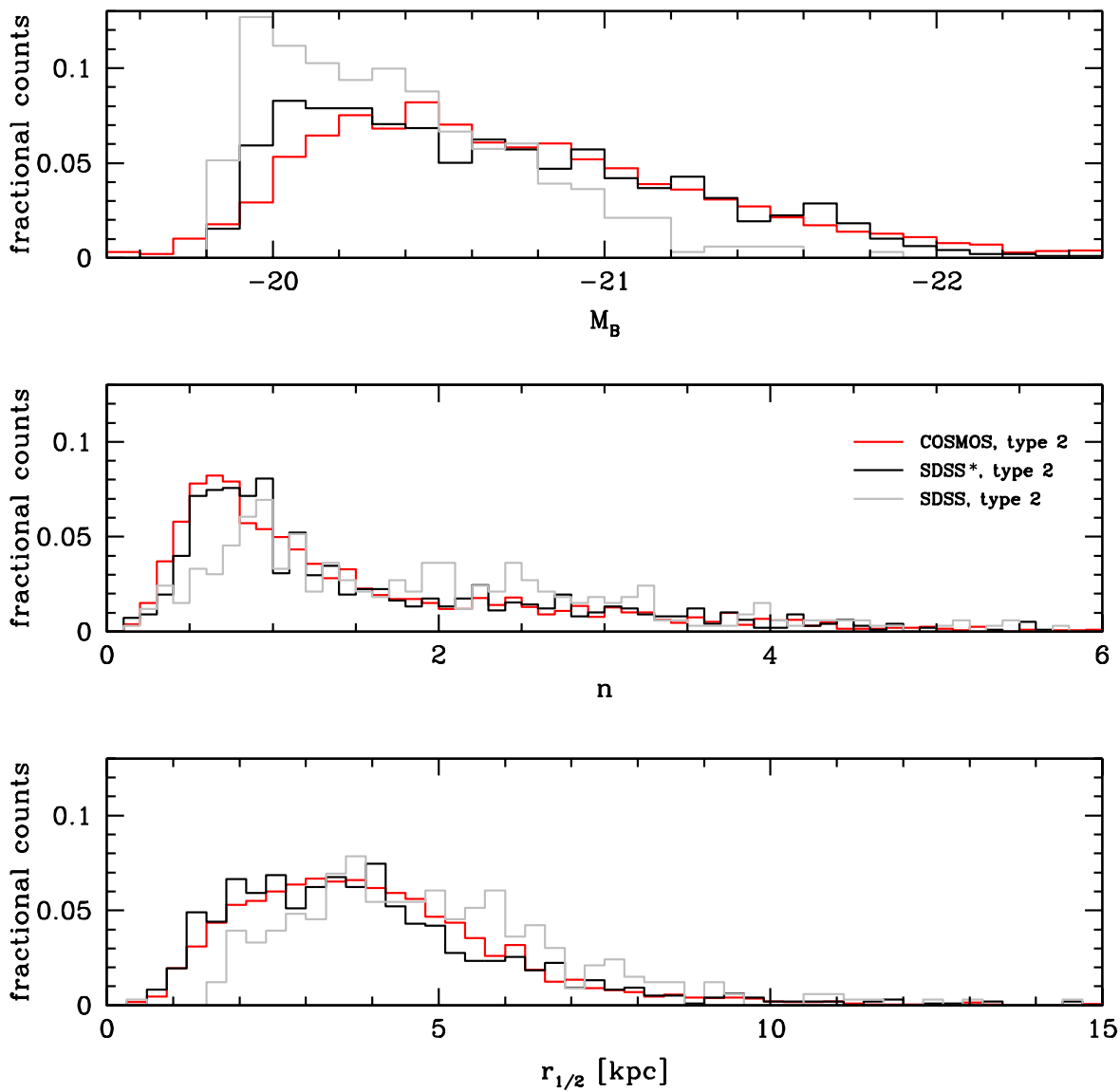}
\caption{\label{sdssdistr} Normalized distribution of absolute B-band magnitudes (top), S\'ersic indices (centre) and half-light radii
(bottom) in the sample of artificially redshifted SDSS galaxies (black histogram; brightened sample - grey histogram; unbrightened sample)
and the COSMOS data set (red histograms) in the range $z\in[0.6,0.8[$.}
\end{figure}

\clearpage

\begin{figure}[ht]
\epsscale{1.0} 
\plotone{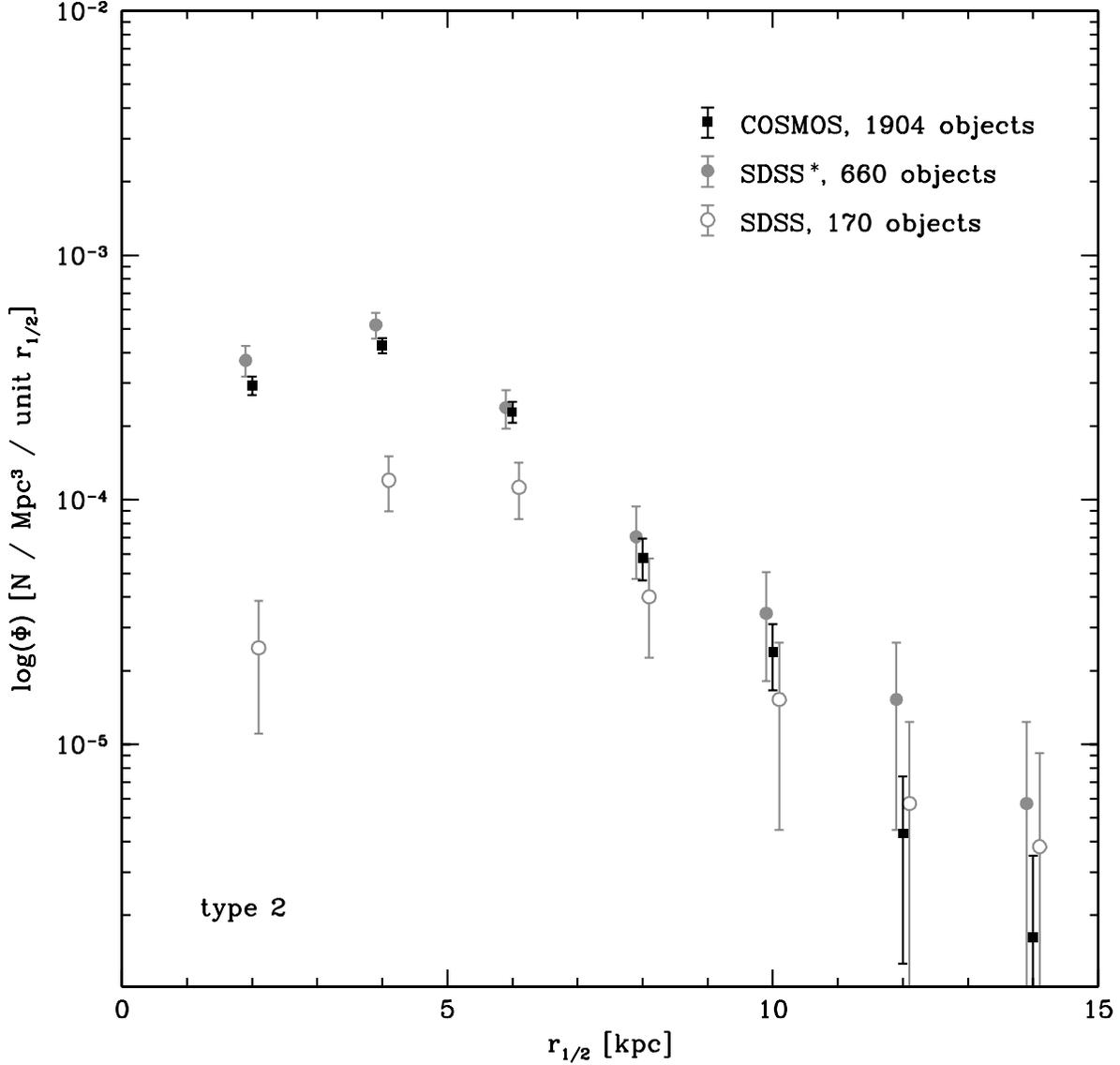}
\caption{\label{sloan1} Size function for COSMOS disk galaxies as plotted in Figure 13 (black squares) in the range $z\in[0.6,0.8[$ and for brightened
(filled grey points) and unbrightened (empty grey points) artificially redshifted SDSS galaxies. The data points of the SDSS size functions have been slightly offset from their nominal value of $r_{1/2}$ for better legibility. Systematic effects on the GIM2D size measurements are identical to those discussed for the COSMOS sample (see caption Figure 13, and Appendix A1). \newline
The COSMOS size function in the range $z\in[0.6,0.8[$ and that of the simulated SDSS$^*$ galaxies closely resemble each other. The steeper slope of the size function in COSMOS relative to that of artificially redshifted SDSS galaxies confirms the tendency observed in Figures 13 and 14, implying a small deficit of the largest disks ($r_{1/2} > 7$ kpc) in the COSMOS sample compared to the local universe.}
\end{figure}

\clearpage

\begin{figure}[ht]
\epsscale{1.0}
\plotone{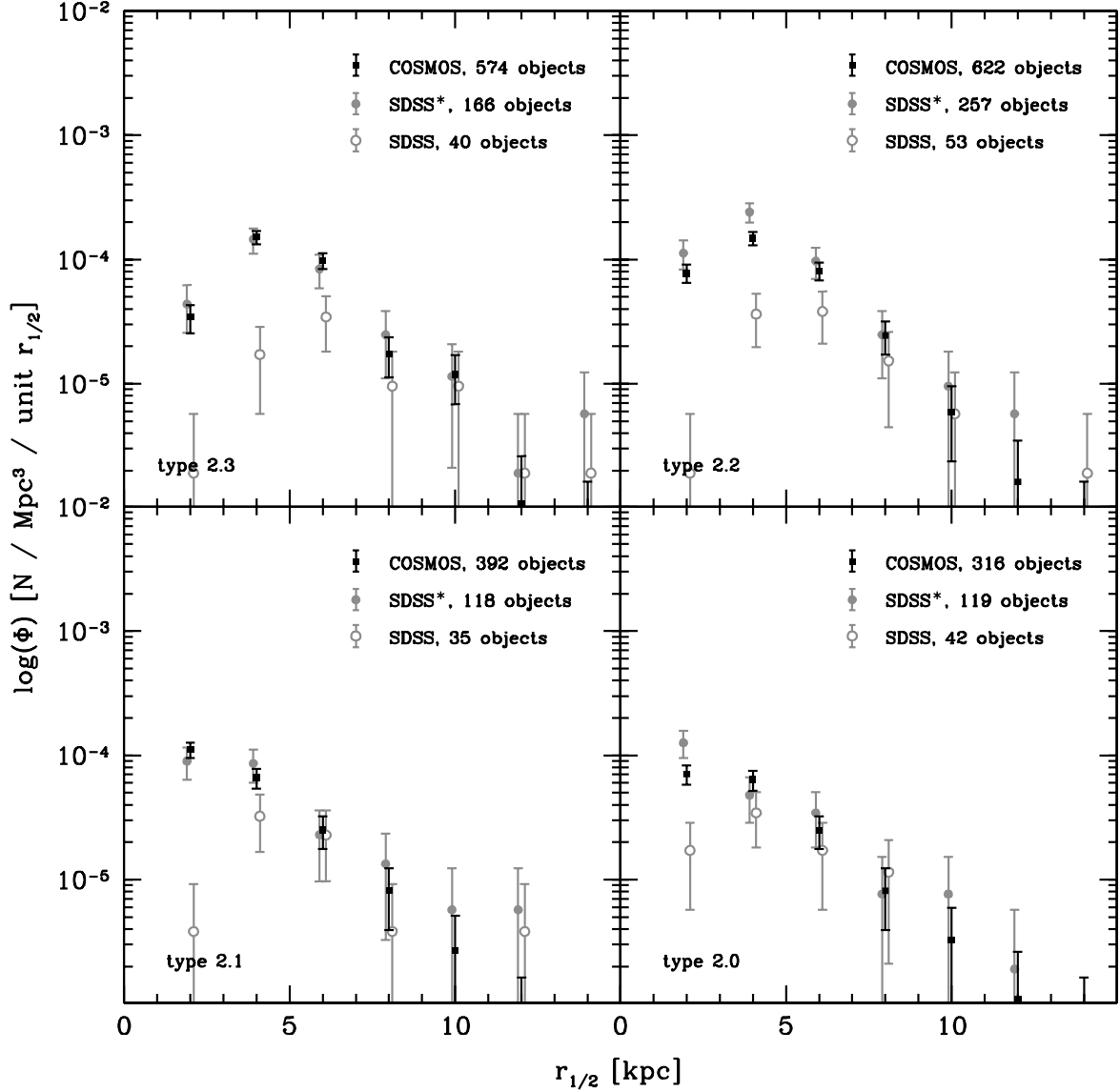}
\caption{\label{sloan2} Size function for COSMOS disk galaxies (reported with black squares and split by relative importance of the bulge component)
in the range $z\in[0.6,0.8[$ (as plotted in Figure 14) and for brightened (filled grey points) and unbrightened (empty grey points) artificially redshifted SDSS galaxies.\newline
Broadly speaking, a general constancy of the size function over half the Hubble time equally is seen for galaxies of all bulge-to-disk ratios. However, the four panels above indicate that the steepening of the slope of the size function is more pronounced for galaxies with measurable bulges than for bulgeless disks. Therefore, the deficit of large disks at earlier epochs is apparently caused by a deficit of {\it bulged} disk galaxies at high redshift.}
\end{figure}

\clearpage

\begin{figure}[ht]
\epsscale{0.8}
\plotone{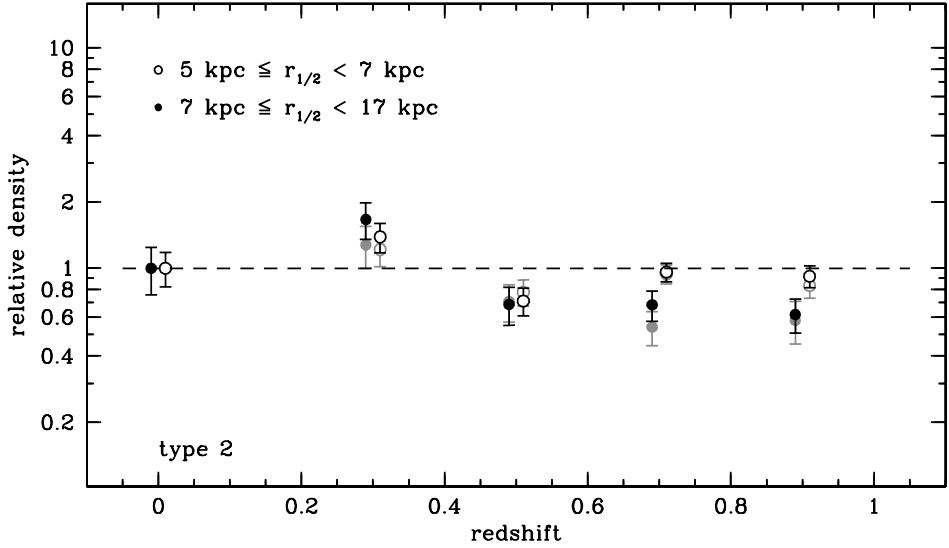}
\caption{\label{sum1} The density of COSMOS disk galaxies with half-light radii in the range 5kpc-7kpc and $>$7kpc, normalized to the corresponding brightened SDSS sample, as a function of redshift (c.f. Table 6 for the values of the relative density $\widetilde{\rho}=\Phi_{COSMOS} /  \Phi_{SDSS^*}$). The point at $z=0$ (equal to 1 by definition
since $\widetilde{\rho}=\Phi_{SDSS^*} /  \Phi_{SDSS^*}$ ) is plotted to show the error on the $z=0$ measurement based on the SDSS sample. The evolution of the SDSS-normalized density is not affected by the systematic errors in the GIM2D estimates of the half-light radii discussed in Appendix A1. Black and grey symbols refer to the two sets of photometric redshifts from ZEBRA, obtained without and with correction to the photometric catalogs (c.f. Section 4.1 and Figures 13 and 14). Uncertainties in the estimates of the photometric redshifts do not change our conclusions. }
\end{figure}

\clearpage

\begin{figure}[ht]
\epsscale{0.8}
\plotone{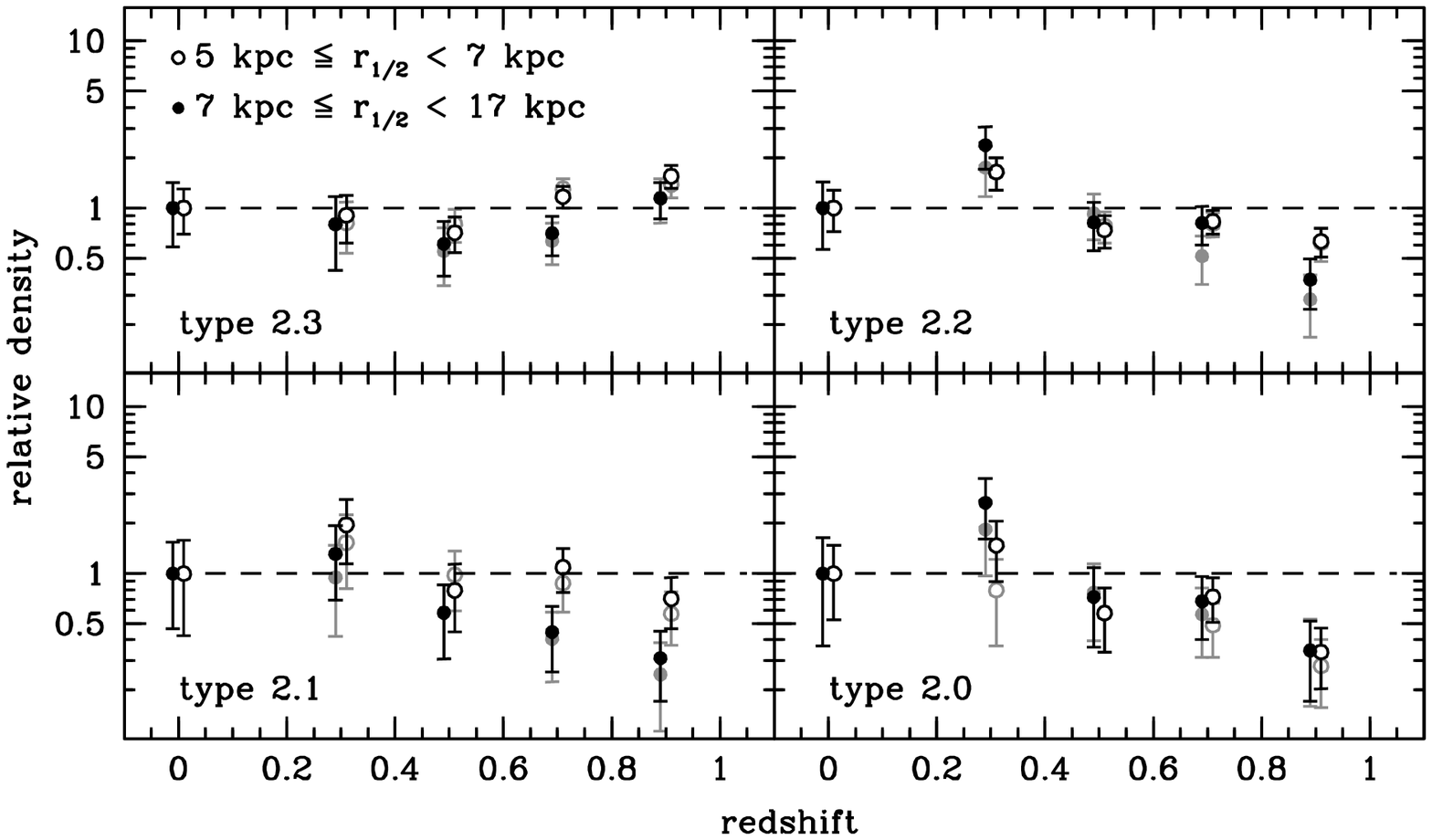}
\caption{\label{sum2} As in Figure 18, but shown individually for disk galaxies with different $B/D$-ratio. Black and grey symbols too are as in
Figure 18.\newline
A constant (or possibly even slightly higher) number density of large bulgeless disks is observed at $z\sim1$, while disk galaxies with a prominent bulge are three time less frequent than in the local universe at that redshift.}
\end{figure}

\clearpage

\begin{figure}[ht]
\epsscale{1.0} 
\plotone{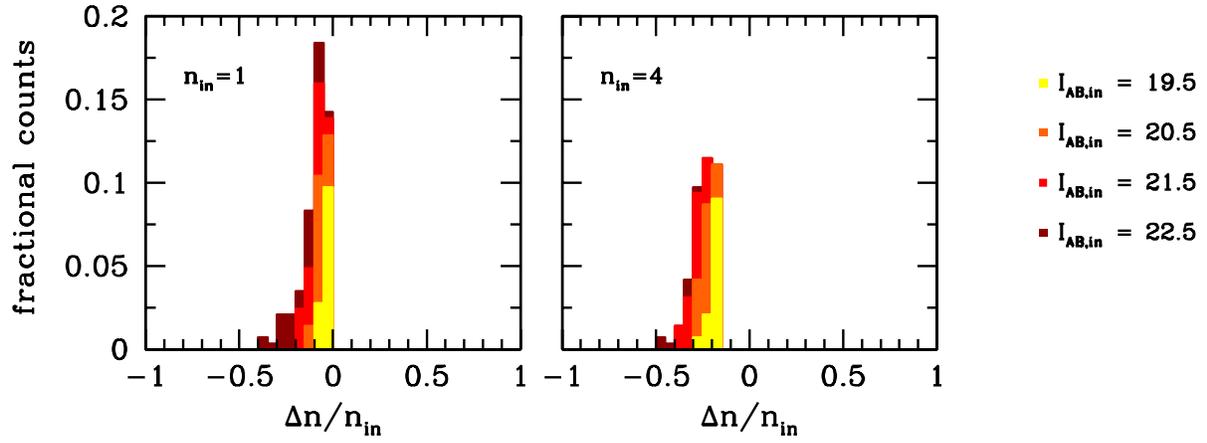}
\caption{\label{simrerr} Dependence of the error in the recovered S\'ersic index $n$ on input magnitude $I_{AB.in}$. On the left we plot the distribution of
normalized difference between the nominal $n$ value and the S\'ersic index recovered by GIM2D for $n_{in}=1$ galaxies, on the right the same for objects
with a nominal S\'ersic index of $n_{in}=4$.}
\end{figure}

\clearpage

\begin{figure}[ht]
\epsscale{1.0} 
\plotone{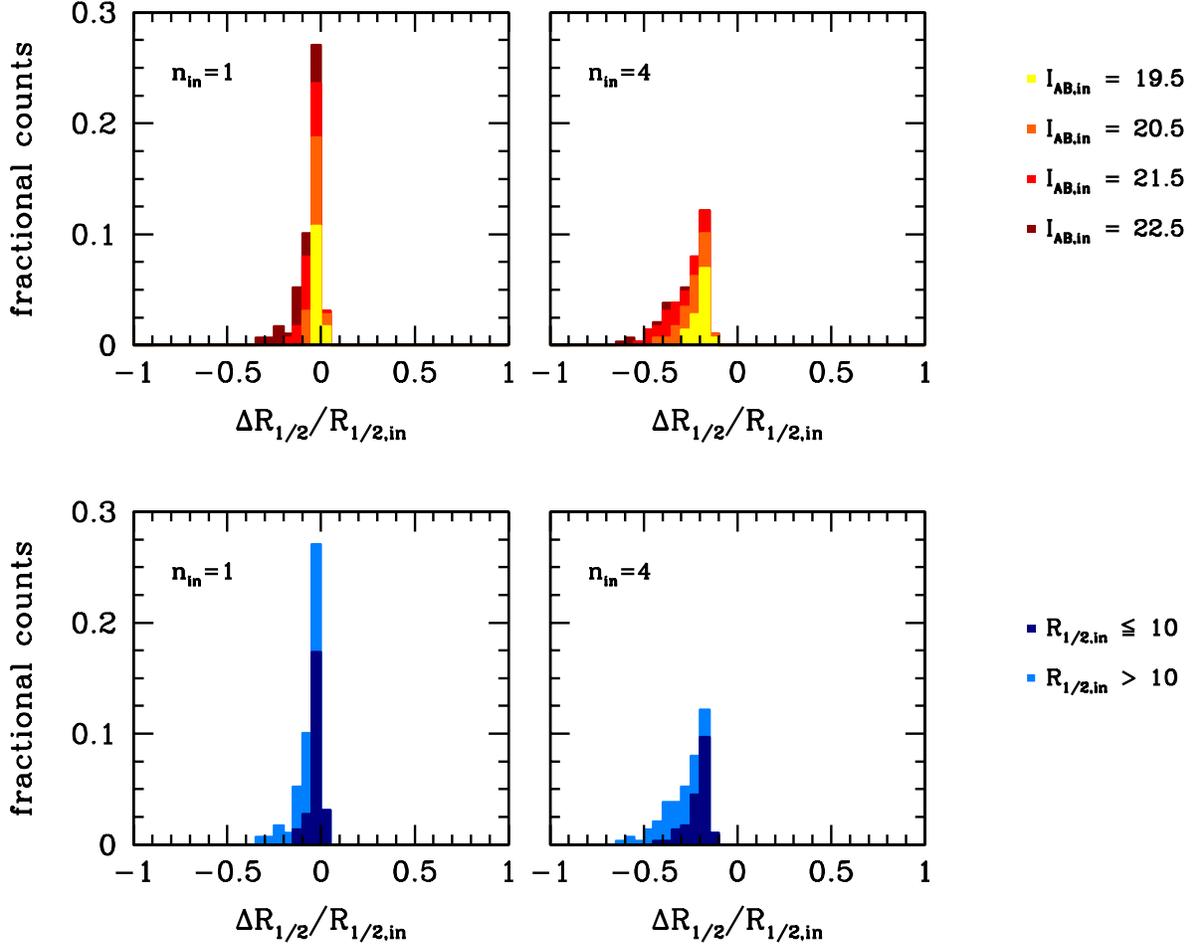}
\caption{\label{simrerr} {\it Upper panel:} Distribution of normalized difference between the nominal input half-light radius of the simulated
galaxies and the half-light radii recovered with the GIM2D fits as a function of magnitude.  {\it Lower panel:} As above, but showing the distribution
for galaxies with input half-light radii smaller and larger than 10 pixels.}
\end{figure}

\clearpage

\begin{figure}[ht]
\epsscale{1.0} 
\plotone{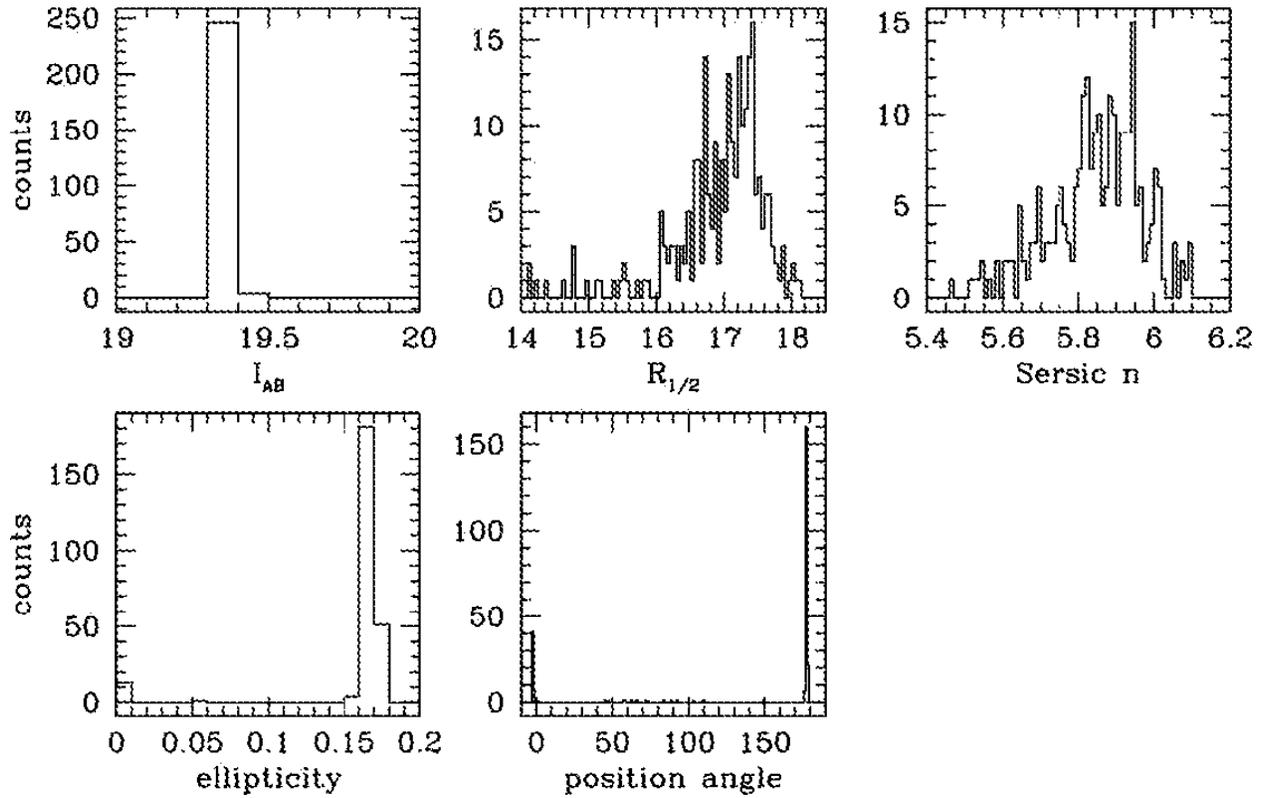}
\caption{\label{psferr} Variation of GIM2D output for a specific galaxy when fit using 250 different point spread functions from the
collection of PSFs employed in our work.  The PSFs have been derived for different individual ACS tiles, and vary as a function of position within a single tile.}
\end{figure}

 \end{document}